\documentclass{emulateapj}
\usepackage{apjfonts}

\begin{document}
\submitted{Accepted for publication in The Astrophysical Journal}

\title{The Dearth of UV-Bright Stars in M32: Implications for
Stellar Evolution Theory\altaffilmark{1}}

\author{Thomas M. Brown, Ed Smith, Henry C. Ferguson}

\affil{Space Telescope Science Institute, 3700 San Martin Drive,
Baltimore, MD 21218;  tbrown@stsci.edu, edsmith@stsci.edu, ferguson@stsci.edu} 

\author{Allen V. Sweigart, Randy A. Kimble, Charles W. Bowers}

\affil{Code 667, NASA Goddard Space Flight Center, Greenbelt, MD
20771; allen.v.sweigart@nasa.gov, randy.a.kimble@nasa.gov, 
charles.w.bowers@nasa.gov}

\altaffiltext{1}{Based on observations made with the NASA/ESA Hubble
Space Telescope, obtained at STScI, and associated with proposal
9053.}

\begin{abstract}

Using the Space Telescope Imaging Spectrograph on the {\it Hubble
Space Telescope}, we have obtained deep far-ultraviolet images of the
compact elliptical galaxy M32.  When combined with earlier
near-ultraviolet images of the same field, these data enable the
construction of an ultraviolet color-magnitude diagram of the hot
horizontal branch (HB) population and other hot stars in late phases
of stellar evolution.  We find few post-asymptotic giant branch (PAGB)
stars in the galaxy, implying that these stars either cross the HR
diagram more rapidly than expected, and/or that they spend a
significant fraction of their time enshrouded in circumstellar
material.  The predicted luminosity gap between the hot HB and its
AGB-Manqu$\acute {\rm e}$ (AGBM) progeny is less pronounced than
expected, especially when compared to evolutionary tracks with
enhanced helium abundances, implying that the presence of hot HB stars
in this metal-rich population is not due to $\Delta Y / \Delta Z
\gtrsim 4$.  Only a small fraction ($\sim$2\%) of the HB population is
hot enough to produce significant UV emission, yet most of the UV
emission in this galaxy comes from the hot HB and AGBM stars, implying
that PAGB stars are not a significant source of UV emission even in
those elliptical galaxies with a weak UV excess.

\end{abstract}

\keywords{galaxies: evolution -- galaxies: stellar content --
galaxies: individual (M32) -- stars: evolution -- stars: horizontal branch }

\section{Introduction}

The nearest elliptical galaxy, M32 (NGC221), provides a useful testing
ground for stellar evolution theory.  With solar-blind ultraviolet (UV)
observations that suppress the dominant cool population, hot stars in
late evolutionary phases can be resolved into the center of the galaxy.
From such observations one can construct a UV color-magnitude diagram (CMD)
for a large sample of hot stars so that these rapid evolutionary phases
can be studied in detail and compared with theoretical predictions.
In contrast, the rapid evolution of these stars
means few of them are found in Galactic globular clusters (the
traditional testing ground for stellar evolution), while distance and
reddening uncertainties hamper their study in the Galactic field
population.

In an elliptical galaxy with little or no star formation, the
UV-bright stars will consist of stars residing on the hot end of the
horizontal branch (HB) itself, also known as the extreme HB (EHB), as
well as stars that have evolved beyond the HB phase.  Although the
HB morphology tends to be red at high
metallicity ($Z$; the ``first parameter''), EHB stars are found in the
Galactic field, metal-rich elliptical galaxies, some of the more massive
metal-rich globular clusters, and metal-rich open clusters.  The existence
of EHB stars in old metal-rich populations demonstrates that
parameters besides metallicity play a role in HB morphology, driving
the ``second parameter'' debate, and producing much controversy
regarding the source of the UV emission (also known as the UV upturn or UV
excess) in quiescent elliptical galaxies.  Both resolved imaging
(Brown et al.\ 2000b) and integrated spectroscopy (Brown et al.\ 1997;
Ferguson et al.\ 1991) imply that EHB stars and their progeny are the
source of this emission, but it is unclear what parameters affect the
large variation in the UV-to-optical flux from galaxy to galaxy (O'Connell 
1999; Yi et al.\ 1999; Tantalo et al.\ 1996; Dorman et al.\ 1995; Park \& Lee
1997; Greggio \& Renzini 1990).  Because a star's effective
temperature on the HB is a function of its mass, possible candidates for
these parameters include age and helium abundance, since both strongly
affect the mass at which a star leaves the main sequence.  

For a population with a given age and metallicity, the stars will arrive
on the HB with nearly the same helium core mass ($\sim$0.5~$M_\odot$)
but a range of hydrogen envelope mass ($\sim$0.001--0.3~$M_\odot$),
determined by the range in mass lost during the ascent up the red
giant branch (RGB).  This range in the envelope mass produces a range
in effective temperature, with the coolest stars having the largest
envelope mass.  After $\sim$100 Myr,
helium is depleted in the convective core, and the star leaves the HB.
Its subsequent evolution to the white dwarf (WD) cooling curve will
occur along one of three possible paths (Figure~1), also depending upon
the envelope mass. Greggio \& Renzini (1990) give a complete review of
these paths, which we briefly summarize here.

\begin{figure}[ht]
\epsscale{1.1}
\plotone{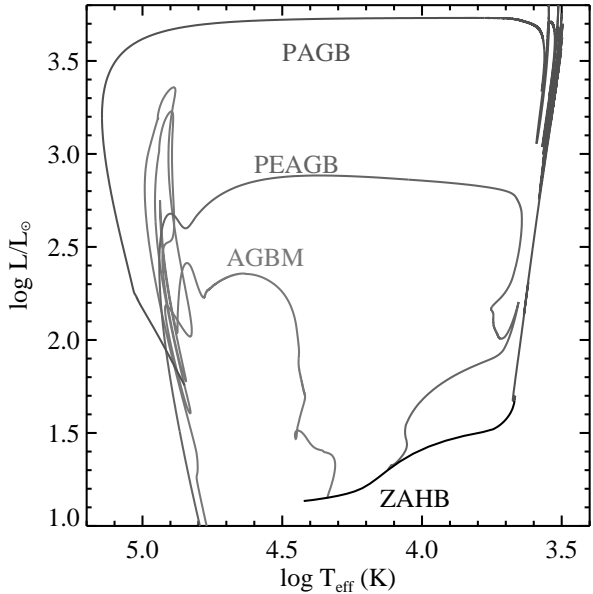} 
\epsscale{1.0}
\caption{Late stages of stellar evolution for solar-metallicity
low-mass stars, shown in the theoretical plane.
Examples of three
possible evolutionary paths from the zero-age HB ({\it black curve;
labeled}) to the WD cooling curve ({\it grey curves; labeled}) are
shown. In an old population, our UV CMD isolates
those evolved stars at $T_{\rm eff} \gtrsim $~10,000~K from the far
more numerous stars on the RGB, AGB, and main sequence.}
\end{figure}

The reddest HB stars (with the most massive envelopes) ascend the asymptotic
giant branch (AGB).  Near the bright end of the AGB, they undergo
thermal pulses, alternating between longer periods of quiescent
hydrogen shell burning and shorter periods of energetic helium shell
flashes.  During the AGB phase, the hydrogen envelope is reduced both by
mass loss (e.g., stellar winds) from the surface and by hydrogen shell burning
at the base of the envelope.  Once the
envelope mass is reduced below a critical value ($M_{crit}$), a star
will leave the AGB and rapidly cross the HR diagram as a bright
post-AGB (PAGB) star, possibly forming a planetary nebula along the
way, before descending the WD cooling curve.  If the
star leaves the AGB between thermal pulses, it crosses as an H-burning
PAGB star, but if it leaves during a thermal pulse, it crosses as an
He-burning PAGB star.  Due to their rapid evolution, PAGB stars are
UV-bright for only a relatively short time (10$^3$--10$^4$ years).  For this
reason most Galactic globular clusters do not currently host any known
hot PAGB stars (e.g., Landsman et al.\ 2000).

HB stars with less massive envelopes ascend the AGB, but their envelope
mass drops below $M_{crit}$ before reaching the thermally-pulsing
stage.  These post-early AGB (PEAGB) stars leave the AGB and cross the
HR diagram at a lower luminosity than the PAGB stars, and then descend
the WD cooling curve.  Because $M_{crit}$ increases
with decreasing luminosity, PEAGB stars will leave
the AGB with a larger envelope than PAGB stars.  This larger envelope,
combined with the lower crossing luminosity, gives the PEAGB stars a
longer UV-bright lifetime between the AGB and WD phases (10$^4$--10$^5$ yr).

EHB stars with very little envelope mass do not ascend the AGB at
all, becoming instead AGB-Manqu$\acute{\rm e}$ (AGBM) or ``failed
AGB'' stars.  They leave the HB and evolve directly to higher luminosities
and higher effective temperatures.  Compared to the PAGB and PEAGB
stars, AGBM stars are UV-bright for a significantly longer time
(10$^6$--10$^7$ yr), which is within an order of magnitude of 
the time spent as an
EHB star ($\sim$ 10$^8$ yr).  During the AGBM phase, most of a star's 
luminosity
comes from the helium-burning shell that forms at the end of the HB
phase.  As the helium-burning shell advances outward through the core,
the temperature within the hydrogen-burning shell increases, until by the
end of the AGBM phase the hydrogen-burning shell begins to make a
non-negligible contribution to the surface luminosity.  The transition
from central helium burning to helium-shell burning at the end of the
HB phase produces a significant luminosity gap between the EHB and AGBM
stars, as will be discussed in section 3.4.

Using the near-UV channel on the Space Telescope Imaging Spectrograph
(STIS; Woodgate et al.\ 1998), Brown et al.\ (2000b) imaged the core of
M32 in order to resolve the source of UV emission in elliptical
galaxies.  The luminosity function of stars in the field demonstrated
that the bulk of the hot evolved population resides on the HB itself.
However, with a single bandpass and a strongly varying bolometric
correction for these evolved stars, it was difficult to disentangle
the various evolutionary phases present in the image.  We subsequently
obtained deep far-UV STIS images of the same field, enabling the
construction of a UV CMD of these populations.
In this paper, we present a detailed comparison of this CMD to the 
expectations from stellar evolutionary theory.

\section{Observations and Data Reduction}

\subsection{HST Observations}

Using STIS on the {\it
Hubble Space Telescope (HST)}, we obtained deep far-UV images
of the compact elliptical galaxy M32.  As done with our earlier STIS
near-UV images (Brown et al.\ 2000b), we targeted a position
$\sim 7.7^{\prime\prime}$ south of the M32 core to allow both the
bright core and the fainter regions of the galaxy to be sampled in
the $25^{\prime\prime} \times 25^{\prime\prime}$ field.  We also
constrained the position angle of the field to maximize the overlap
with our earlier near-UV images.  Unfortunately, this overlap
was somewhat reduced by an intrinsic offset between the field centers
defined for the two STIS UV channels, such that the region with at
least 80\% exposure depth in both channels is approximately
$21^{\prime\prime} \times 23^{\prime\prime}$ in size.  Our analysis
here is restricted to this region.

\begin{table}
\begin{center}
\caption{Observations}
\begin{tabular}{rrrc}
\tableline
            & \multicolumn{2}{c}{Exposure Time} &  \\
Date        & Total & Low Sky & Position Angle \\
\tableline
30 Jul 2001 & 13472 s  &  9640 s  & -157$^{\rm o}$           \\
31 Jul 2001 & 13472 s  &  9640 s  & -157$^{\rm o}$           \\
 2 Aug 2001 & 13472 s  &  9340 s  & -157$^{\rm o}$           \\
 4 Aug 2001 & 13472 s  &  9040 s  & -157$^{\rm o}$           \\
12 Oct 2002 & 13472 s  & 10560 s  & +113$^{\rm o}$           \\
13 Oct 2002 & 13472 s  & 10560 s  & +117$^{\rm o}$           \\
22 Oct 2002 & 13472 s  & 10560 s  & +117$^{\rm o}$           \\
\tableline
\end{tabular}
\end{center}
\end{table}

Our total exposure time in the strontium fluoride filter (F25SRF2) was
94,304 s, split into 7 visits of 5 orbits each (Table 1).  The F25SRF2
filter blocks geocoronal emission from Lyman $\alpha$ but not
\ion{O}{1} $\lambda 1301$ or \ion{N}{1} $\lambda 1493$ (Brown et al.\ 2000c).  
The \ion{O}{1} emission varies strongly over an orbit, so
we split each visit into a series of 47 exposures of 300 s each (with
some of these exposure times adjusted to accommodate orbit boundaries).
We did not use the STIS TIME-TAG mode, where the exposure records the
location and time for every detected photon, because this mode
requires careful management of the data buffer, which is difficult to
do when the sky is varying so strongly.  Because the STIS UV detectors
are photon-counting multianode microchannel arrays (MAMAs), there is
no read noise.  Furthermore, the instrument can transfer data from its
internal buffer to the {\it HST} data recorder during an exposure if
that exposure is at least 300 s.  For these reasons, there was no
penalty in obtaining a series of 300 s exposures instead of one long
exposure per orbit, and the advantage of this approach was the ability
to remove periods of high sky background due to geocoronal \ion{O}{1}
emission.  Note that the MAMA detectors also register less than one
count per incident cosmic-ray, so cosmic-ray rejection is
not required, and breaking up an orbit into multiple exposures is not
a concern, as it is for CCDs on {\it HST} (where one generally wants
to keep exposures to $\sim$1000 s or less). \\

\subsection{Data Reduction}

After processing the far-UV data through the standard CALSTIS
pipeline, we inspected a region free of stars in each 300 s exposure
in order to cull those exposures with high sky background.  Our
threshold was 0.03 counts s$^{-1}$ pix$^{-1}$, with periods of low sky
generally exhibiting count rates of $\sim$0.01 counts s$^{-1}$
pix$^{-1}$ and periods of high sky exhibiting count rates
10 to 100 times higher.  We then coadded the low sky exposures within
each visit, because they were all taken at the same position (i.e.,
there was no dithering during a visit).  The exposure times retained
for each visit are listed in Table 1.  We then applied a geometric
distortion correction to each visit's image, using the IRAF DRIZZLE
package (Fruchter \& Hook 2002).  Because of inaccuracies in the STIS
geometric distortion solution, the residual distortions in the
resulting images did not allow satisfactory co-addition of the 2001
data with the 2002 data, given the $\sim$90$^{\rm o}$ roll between
those datasets (see Table 1).  Furthermore, the far-UV images could
not be adequately registered with our earlier near-UV image (Brown et
al.\ 2000b).  Thus, using the IRAF GEOMAP task, we derived an
additional distortion solution to transform the far-UV images to the
near-UV reference frame, utilizing several hundred bright stars that
could be identified as common to both bandpasses.  We then used the
IRAF GEOTRAN task to transform the far-UV images, and then coadded the
far-UV data to a single image (Figure~2).

\begin{figure*}[ht]
\epsscale{1.1}
\plotone{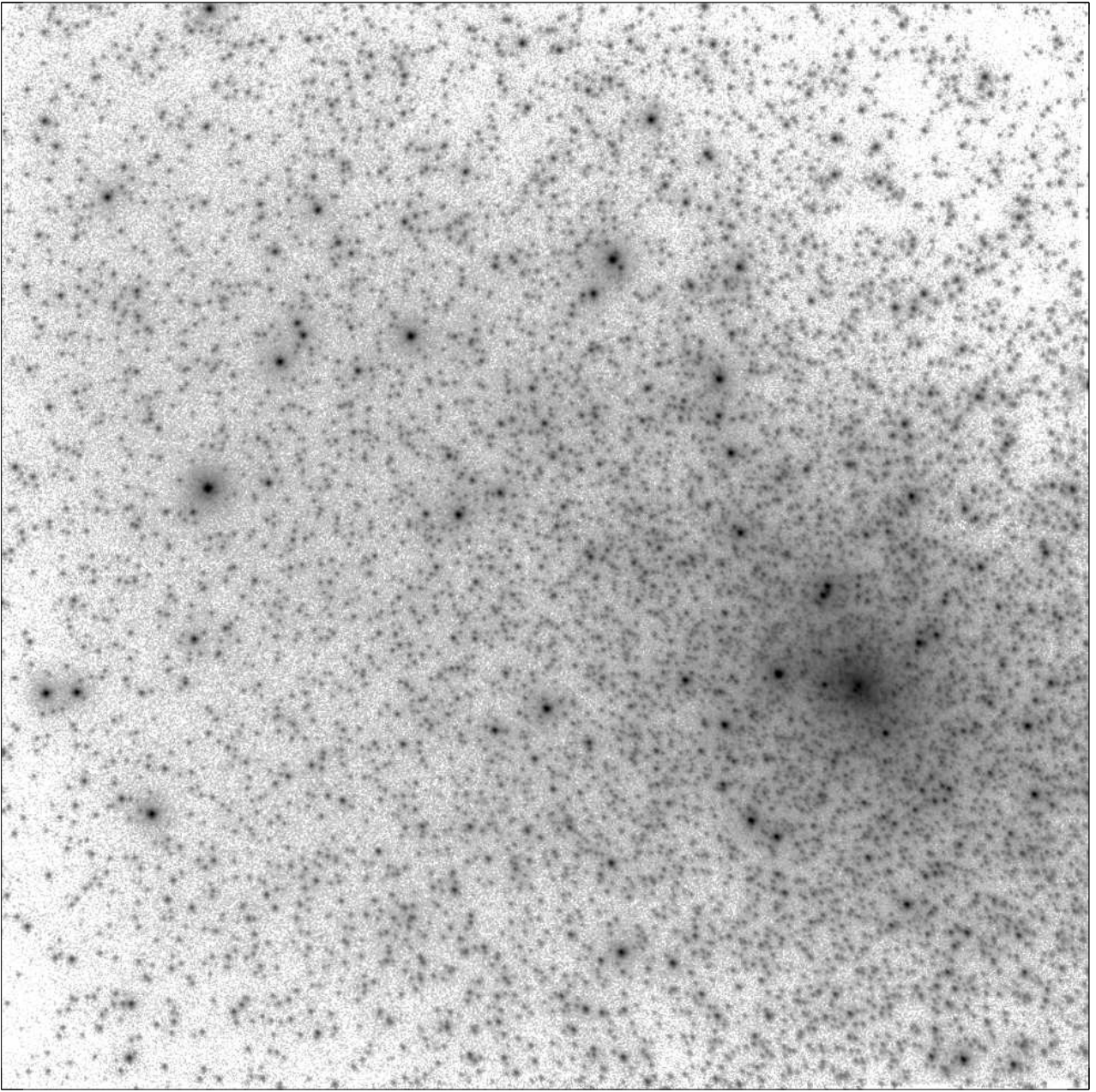} 
\epsscale{1.0}
\caption{The STIS/FUV image of M32, with a logarithmic stretch and cropped
to 24$^{\prime\prime} \times 24^{\prime\prime}$.}
\end{figure*}

We used the DAOPHOT-II package (Stetson 1987) to construct a model
point spread function (PSF) from isolated stars in the far-UV image.
We then performed PSF-fitting photometry on the far-UV image, using
two passes of object detection and fitting.  Matching this photometry
to the near-UV photometry of the same field (Brown et al.\ 2000b)
proved to be difficult for two reasons.  Due to residual small-scale
differences in the geometric distortion and the variation in PSF shape
with position, the inconsistency between the two bandpasses for a
star's position significantly varied across the field, with the offset
in each axis having an rms of $\sim$0.3~pixels; the effects
of this mis-registration was included in our analysis during the
artificial star tests.  Furthermore, the bandpasses are very distinct
in wavelength, such that a star could be much brighter than its
neighbors in one bandpass but much fainter than its neighbors in the
other, to the extent that stars detected in one bandpass were often
undetected in the other.  In the end, we decided to perform aperture
photometry (2 pixel radius) in the near-UV image at the positions of
objects detected and fit in the far-UV image.  Although the catalog of
Brown et al.\ (2000b) used PSF-fitting photometry in the near-UV
image, we did not employ PSF-fitting in the near-UV photometry for the
current analysis; fixing the near-UV PSF fitting at the far-UV
positions caused fits to fail, due to the mis-registrations, while
allowing the near-UV PSF fitting to float in position frequently
caused the fits to wander onto neighboring stars, due to both the
mis-registration and the large color variations from star to star.
The final catalog was cleaned of severe blends by using the quality of
the PSF fit in the far-UV photometry and a concentration index (i.e.,
comparing photometry in apertures of two distinct radii) in the
near-UV photometry.  The catalog was also cleaned of stars falling
within the bright center of the galaxy (an irregularly-shaped region
of area 9.4 arcsec$^2$).  Although we obtained PSF-fitting photometry
of 7,500 stars in the far-UV image, only 5,200 of these resulted in
legitimate near-UV photometry and survived the catalog cleaning.

The photometry for each bandpass was put on an absolute magnitude
scale by normalizing to aperture photometry on the brightest stars.
For the near-UV photometry, this was done by comparing aperture
photometry for radii of 2 and 3 pixels and then using the aperture
correction of Proffitt (2003) to go from 3 pixels to infinity; for the
far-UV photometry, this was done by comparing the PSF-fitting
photometry to that in a 3 pixel aperture, and then applying the
correction of Proffitt (2003) to go from 3 pixels to infinity.  Our
photometry is in the STMAG system: $m = -2.5$~log$_{10} f_\lambda
-21.1$~mag, where $f_\lambda = e^-$(PHOTFLAM/EXPTIME), EXPTIME is the
exposure time, and PHOTFLAM is $5.836 \times 10^{-18}$ erg s$^{-1}$
cm$^{-2}$ \AA$^{-1}$ ($e^-$ s$^{-1}$)$^{-1}$ for the near-UV bandpass and 
$4.201 \times 10^{-17}$ erg s$^{-1}$ cm$^{-2}$ \AA$^{-1}$ 
($e^-$ s$^{-1}$)$^{-1}$
for the far-UV bandpass.  These PHOTFLAM values take into account the
time-dependent sensitivity of the STIS UV bandpasses, and represent
the exposure-time weighted average for the dates of the observations.
The STMAG system is a convenient system because it is referenced to an
unambiguous flat $f_\lambda$ spectrum; an object with $f_\lambda =
3.63 \times 10^{-9}$ erg s$^{-1}$ cm$^{-2}$ \AA$^{-1}$ has a magnitude
of 0 in every bandpass.

\begin{figure}[ht]
\epsscale{1.1}
\plotone{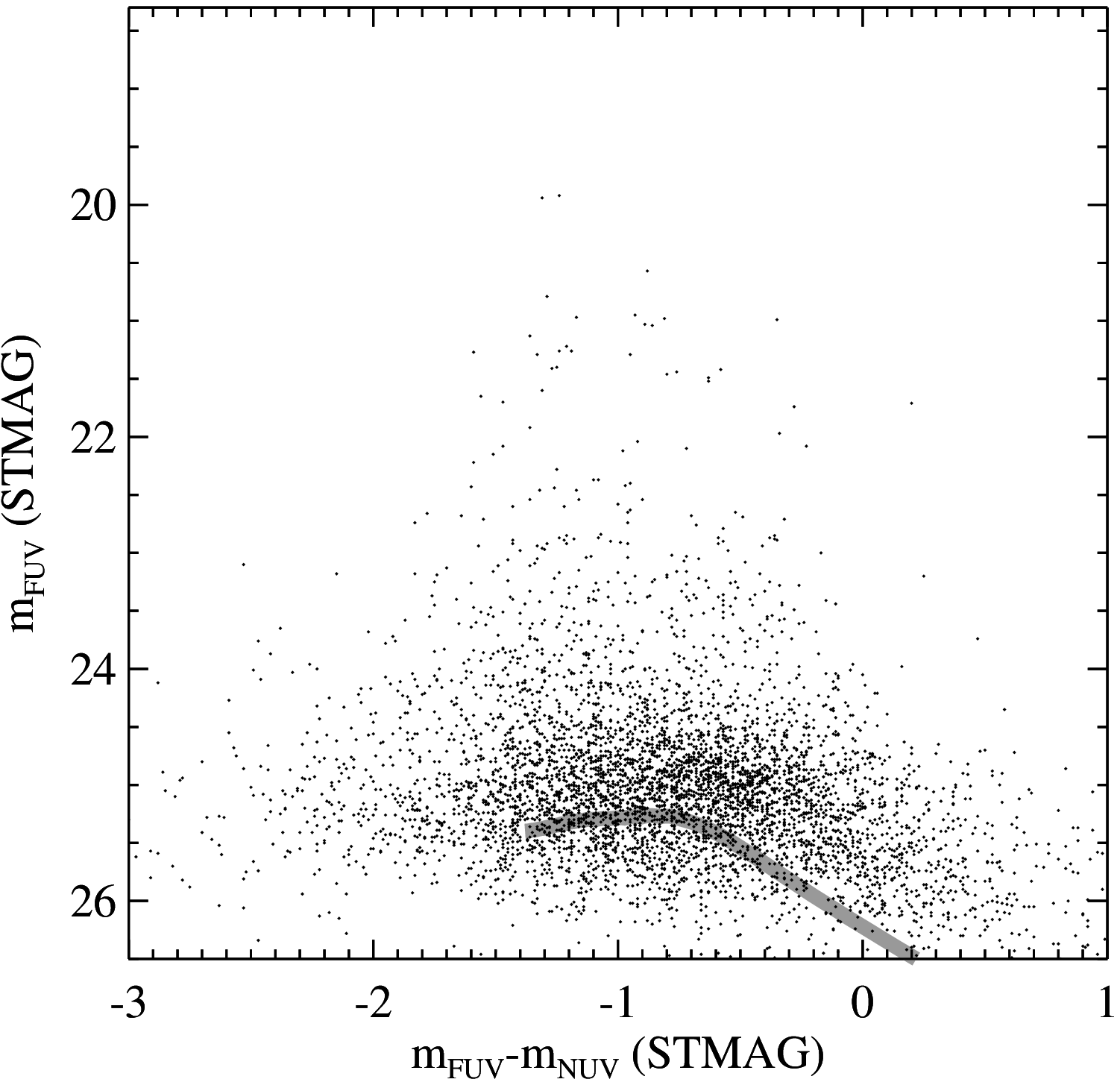} 
\epsscale{1.0}
\caption{The UV CMD of M32.  The EHB and UV-bright post-HB stars are
clearly resolved.  A solar-metallicity zero-age HB is plotted for
reference ({\it grey curve}).
}
\end{figure}

Artificial star tests were performed by inserting 400 artificial stars
into the near-UV and far-UV images, blindly recovering them with the
same process used to create our photometric catalog, and then
repeating this process 12,500 times (for a total of 5,000,000
artificial stars).  The artificial stars spanned the range $-2.5 \leq
(m_{FUV}-m_{NUV}) \leq 1$~mag and $19.5 \leq m_{FUV} \leq 27$~mag.  No
stars brighter than this range were found in the data; in our modeling, 
those few stars falling in brief evolutionary phases brighter than this
range were assumed to have the same small scatter and nearly 100\%
completeness as the bright end of the artificial star tests.  Although
the photometric process uses aperture photometry for the near-UV
measurements, a spatially-variable model of the near-UV PSF
(constructed from isolated stars in the near-UV data) was used to
insert artificial stars into the near-UV image.  Artificial stars were
inserted into the near-UV image with a small random offset from the
position in the far-UV image ($\sigma$~=~0.3 pixels in each axis), to
approximate the effects of the small-scale residuals in the geometric
distortion correction (discussed above).

Artificial stars were not distributed uniformly across the far-UV and
near-UV images.  Instead, because object detection in our process is
done on the far-UV image, the density of stars in each pass of the
artificial star tests was matched to the far-UV surface-brightness
profile, created by applying a ring median filter to the far-UV image.
This spatially non-uniform distribution of artificial stars ensures
that the resulting photometric scatter and incompleteness matrix is
properly weighted to the population we are trying to model.  To
understand the importance of this weighting, it is useful to consider
a simplified example.  

Imagine a galaxy with an abrupt edge, imaged so
that this edge bisects the camera's field of view.  In the left-hand
half of the image, there is no significant luminosity from the galaxy,
and in the right-hand half of the image, there is a crowded field of
stars.  Assume that in the empty left-hand side of the image, an
isolated EHB star could in principle be detected at a completeness
of nearly 100\% and measured with a photometric error of 0.02~mag,
while in the crowded right-hand half of the image, such a star could
be detected at 50\% completeness and measured with a photometric error
of 0.2~mag.  Also assume that much brighter PAGB stars could be
detected at nearly 100\% completeness and measured with $\sim$0.01~mag
photometric errors anywhere in the image.  If one performed artificial
star tests on such an image with a uniform distribution of stars,
those tests would give an erroneous completeness correction and imply
an erroneous ratio of EHB to PAGB stars in the field.  Specifically,
one would incorrectly find a completeness of 75\% and a photometric
error of 0.08~mag for the EHB stars.  In reality, the population being
studied only exists in the right-hand side of the image, where EHB
stars have lower completeness and larger photometric errors.  Instead,
by distributing the artificial stars so that their density follows the
luminosity in the image, one recovers an accurate measure of the
completeness and photometric scatter in the entire population.
Alternatively, one could uniformly distribute the artificial star
tests and weight the results by the luminosity profile in the image.

The knowledge of the completeness and photometric scatter can then be
applied to the modeling in different ways.  One can attempt to correct
the data before comparison to the models, but it is difficult to
deconvolve the photometric scatter.  Instead, in our analysis here, we
work in the opposite sense: we apply the incompleteness and
photometric scatter to the models to create simulations, and then
compare to the uncorrected data.

\subsection{The Color-Magnitude Diagram}

We show the UV CMD of M32 in Figure~3.  The most obvious feature of
the CMD is the presence of a hot HB population.  The existence of these
stars was inferred by Brown et al.\ (2000b) from the luminosity
function in a single near-UV bandpass; with the present CMD, the
detection of an EHB population in M32 is unequivocal.  The other
striking features of this CMD are the small number of UV-bright stars
above the HB and the lack of a luminosity gap between the EHB and
these UV-bright stars.  We will explore these aspects of the CMD in
the modeling below.  Note that in the region of M32 spanned by our UV
catalog (i.e., the sky sampled by both the near-UV and far-UV images
but excluding the 9.4 arcsec$^2$ region in the M32 core), the optical
surface brightness of M31 is $<<$1\% that of M32 (Walterbos \&
Kennicutt 1987), so contamination from M31 stars should be
negligible.

\section{Evolutionary Sequences}

\subsection{HB and post-HB Tracks}

We have calculated a grid of HB and post-HB evolutionary tracks with a
variety of abundances appropriate for analyzing the UV CMD of
M32.  These tracks span a wide range of HB masses from the hot end of the
EHB to the red HB with a fine mass spacing, in order to capture
the full range of post-HB behavior.  For most tracks, the evolution
was followed through the HB phase using standard algorithms
for convective overshooting and semiconvection (Robertson \& Faulkner
1972) and then through the post-HB phase
until the models reached a faint luminosity on the WD cooling
curve.  However, a small subset of our models encountered a final
helium-shell flash while
descending the WD cooling curve.  The calculations for these
models were stopped if the resulting flash convection extended into
the hydrogen envelope.

For each composition we first obtained a zero-age HB (ZAHB) model at the
red end of the HB by evolving a model from the zero-age main sequence,
up the RGB, and then through the helium-core flash at the tip of the
RGB.  The initial mass for these sequences was adjusted for each
composition to give an age of 13 Gyr at the ZAHB phase.  This assumed
age for the HB stars in M32 is not, however, important for our analysis,
since the helium-core mass of a ZAHB model is only weakly dependent
on the initial mass of the star.  Assuming a younger age, and hence larger
main sequence mass, would simply yield a slightly redder ZAHB model with
a larger envelope mass but the same helium-core mass.  We then produced
the hotter ZAHB models needed for our HB and post-HB tracks by removing mass
from the envelope of our red ZAHB model.

We computed evolutionary tracks for three values of the metallicity:
[Fe/H]~=~0.0, $-0.25$ and $-0.45$, corresponding to heavy-element
abundances $Z$ = 0.01716, 0.01, and 0.006, respectively.  Scaled solar 
abundances
were assumed for these metallicities with no enhancement of the $\alpha$
elements, i.e., [$\alpha$/Fe]~=~0.0.  These metallicities cover the range
spanned by the bulk of the M32 population (see Worthey et al.\ 2004), with
[Fe/H]~=~$-0.25$ lying near the peak of the M32 metallicity
distribution.  The
tracks with [Fe/H]~=~0.0 were calibrated by adjusting the
heavy-element abundance $Z$ to 0.01716, the helium abundance $Y$ to
0.2798, and the mixing-length ratio $\alpha$ to 1.8452, in order to
match the observed solar luminosity, radius, and $Z/X$ ratio at an age
of 4.6~Gyr.  The helium abundance of these solar metallicity models increased
to 0.3003 during the first dredge-up on the RGB, and this therefore represents
the envelope helium abundance during the subsequent HB and post-HB evolution.
The main-sequence helium abundance of the [Fe/H]~=~$-0.25$ and $-0.45$ 
sequences was set at $Y$~=~0.23, which increased to 0.2518 and 0.2496, 
respectively, during
the first dredge-up.  We also calculated a set of EHB models
(including their post-HB progeny) with enhanced helium abundances
of $Y$~=~0.28, 0.33, 0.38, and 0.43 in order to explore the strong effect
of $Y$ on the HB and post-HB evolution.

Mass loss was included during the AGB phase using the Reimers
(1975) formalism, which is parameterized by the mass-loss parameter
$\eta_R$.  We calculated AGB tracks for $\eta_R$ values of both 0.4 and 1.0.
In \S3.3, we will show that both the lifetime and mean luminosity of
a PAGB star within the UV-bright region of Figure~3 depend primarily on the 
final
mass of the star.  Thus a bluer HB star that evolves up the AGB with a smaller
mass loss can have the same UV-bright lifetime and mean luminosity as a redder 
HB star that evolves with a larger mass loss if the final PAGB masses are the
same.  Our use of different $\eta_R$ values was simply a means for producing
PAGB stars over a range in mass.  It was not intended to represent the actual
mass loss process along the AGB.  Finally we note that there was no mass loss
in the AGBM sequences, which do not ascend the AGB at all.

\subsection{Translation to Observed Parameters}

We translate our HB and post-HB tracks (and others from the literature) into
observed magnitudes ($m_{FUV}$ and $m_{NUV}$) by folding synthetic
spectra through the STIS effective area curves.  We use the Castelli
\& Kurucz (2003) grid of synthetic spectra for $T_{\rm eff} \leq
50,000$~K and blackbody spectra for hotter stars.  The grid of
synthetic spectra is interpolated in effective temperature, surface
gravity, and metallicity, assuming scaled solar abundances (no alpha
enhancement).  Because the synthetic spectra cannot account for any
helium enhancement, we only match the spectra to the $Z$ value of
our evolutionary tracks, even when $Y$ is enhanced.  We
assume that M32 is at the same distance as M31 (770 kpc; Freedman \&
Madore 1991), and we generally assume a foreground extinction of $E(B-V)
= 0.08$~mag (Schlegel et al.\ 1998) and the mean Galactic extinction
law of Fitzpatrick (1999).  In addition, we also consider a small dispersion in
reddening for the sightlines to the stars in our catalog.  
Potential sources of reddening variation in the field include
Galactic foreground, M31 disk (if M32 lies slightly behind the M31
disk), and dust intrinsic to M32 itself (which should be minimal; see
Corbin et al.\ 2001).  At this time, it is unclear if M32 lies behind
the M31 disk (see Worthey et al.\ 2004) or in front of it (see
Ford et al.\ 1978; Paulin-Henriksson et al.\ 2002).  

\subsection{Post-HB Evolution}

As described in \S1, the post-HB evolutionary path of a star is driven
by its envelope mass on the HB.  At the one extreme, red HB stars with
large envelopes give rise to PAGB stars, while at the other extreme,
EHB stars with small envelopes give rise to AGBM stars.  Regardless of
the evolutionary path, all stars leaving the HB will spend at least some
time as a hot post-HB star within the UV-bright region of Figure~3.  For
the purposes of the discussion here,
we define a ``hot post-HB star'' as a star residing at
$(m_{FUV}-m_{NUV}) < 0$~mag and 0.5~mag brighter than the end of the
core helium-burning phase for a solar-metallicity HB (see Figure~4).
The luminosity and duration of this hot post-HB phase
depend on a star's final mass on the WD cooling curve.
A star with high mass will evolve through the PAGB phase at high luminosities
on a short timescale (thousands of years), while a star with low mass will
evolve through the AGBM phase at low luminosity on a long timescale
(millions of years).

In order to explore these properties in more detail, we have calculated both
the time spent within the ``hot post-HB star'' region of Figure~4 and the mean
bolometric luminosity within this region for each of our post-HB evolutionary
tracks.  The results for our solar-metallicity tracks,
given in Figure~5, show a tight correlation from
the faint, but long-lived, AGBM tracks to the bright, but short-lived,
PAGB tracks.  Our tracks with AGB mass-loss rates of $\eta_R$~=~0.4
and $\eta_R$~=~1.0 fall on the same trend, indicating that the correlation
in Figure~5 is insensitive to the details of the AGB mass-loss process.
For comparison, we also plot the same results for the solar-metallicity
tracks of Dorman et al.\ (1993; hereafter DRO93), Yi et al.\ (1997;
hereafter YDK97), and Vassiliadis \& Wood
(1994; hereafter VW94), and in all of these cases we find good agreement with
our models.  The H-burning and He-burning PAGB tracks of VW94
in Figure~5 correspond to main sequence turnoff masses of 1.0, 1.5, and
2.0 $M_\odot$.  With the mass-loss assumptions of VW94, these
main-sequence masses produce PAGB stars having
masses of 0.57 -- 0.63~$M_\odot$.  The 0.633 $M_\odot$ H-burning
PAGB track of VW94 that evolved from a 2 $M_\odot$ main-sequence
star is similar to our own $\sim$0.6 $M_\odot$ tracks that evolved
from a 1 $M_\odot$ main-sequence star, further suggesting that
the properties of a PAGB star are mainly governed by the star's final
mass.

Because variations in metallicity will be one of the parameters that we 
will later explore in \S4 when trying
to model the M32 UV CMD, it is worth determining whether the correlation shown
in Figure~5 depends upon metallicity.  In Figure~6, we show a similar
plot, but limited to our own tracks at three values of [Fe/H]: +0.0 (solar),
$-0.25$, and $-0.45$.  The same tight correlation between 
lifetime and luminosity found in Figure~5 persists for all of these
metallicities.

Figure~5 strongly indicates, but does not prove, that the duration
of the hot post-HB phase and its time-averaged bolometric luminosity
are each determined by a star's final mass.  In order to examine
this point more closely, we plot each of these quantities separately in 
Figure~7 as a function of the final mass for both our tracks and those
of DRO93, YDK97, and VW94.  Again we see that the long-lived AGBM tracks
are clearly separated from the shorter-lived PEAGB and PAGB tracks.
The DRO93 tracks
deviate towards shorter durations but brighter luminosities for the
hot post-HB phase.  Except for this deviation all of the other tracks
in Figure~7 agree well.  We conclude that the properties of our tracks
within the hot post-HB star region are primarily determined by the final mass
and are not dependent on the details of the prior mass loss on the
AGB.

In order to utilize our post-HB evolutionary tracks in the analysis of
the UV CMD of M32,
we need to replace the theoretical parameters in Figure~5 with the
corresponding observable parameters.  Replacing the bolometric
luminosity with the far-UV magnitude is straightforward, as discussed
in \S3.2.  However, replacing the duration of the hot post-HB phase with
the expected number of hot post-HB stars for a given evolutionary
track requires a more detailed explanation.

\begin{figure}[ht]
\epsscale{1.2}
\plotone{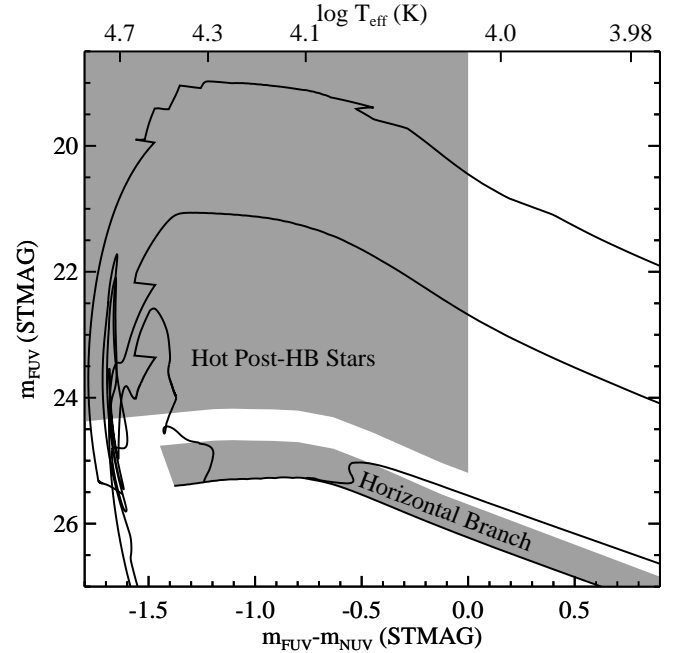} 
\epsscale{1.0}
\caption{The evolutionary paths shown in Figure~1, but transformed
to the STIS UV bandpasses assuming a
distance of 770~kpc (Freedman \& Madore 1991) and $E(B-V) = 0.08$~mag
(Schlegel et al.\ 1998).  The 
region used to define the ``hot post-HB stars'' in the present discussion
is highlighted ({\it grey shading; labeled}).
The relation between $T_{\rm eff}$ and color on the x-axis
assumes log~$g$~=~5 and [Fe/H]~=~0.  
}
\end{figure}

\begin{figure}[ht]
\epsscale{1.2}
\plotone{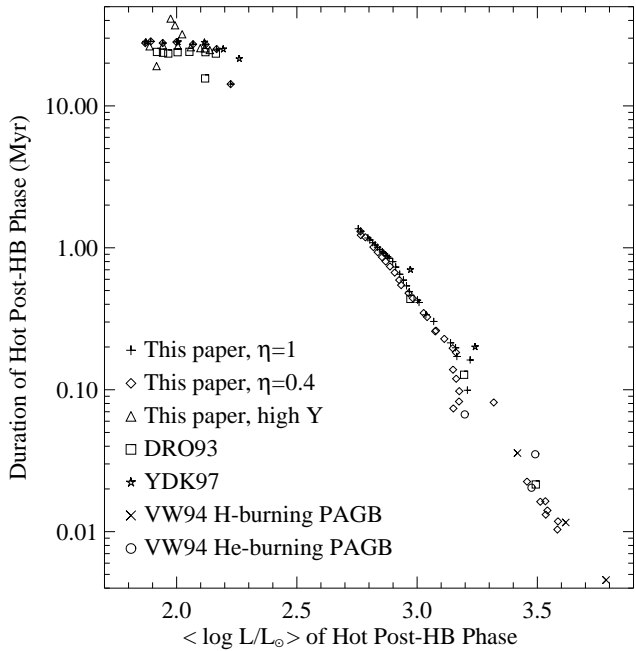} 
\epsscale{1.0}
\caption{The time spent as a hot post-HB star as a function of the
time-averaged luminosity in this phase.  The AGBM stars are
relatively long-lived and faint, while the PAGB stars are relatively
short-lived and bright.  The gap in the distribution at 
log $L \approx 2.5$~$L_{\odot}$
separates the AGBM stars from the PEAGB stars.
Other evolutionary tracks from the literature
are also shown, giving good agreement with our own tracks.}
\end{figure}

\begin{figure}[ht]
\epsscale{1.2}
\plotone{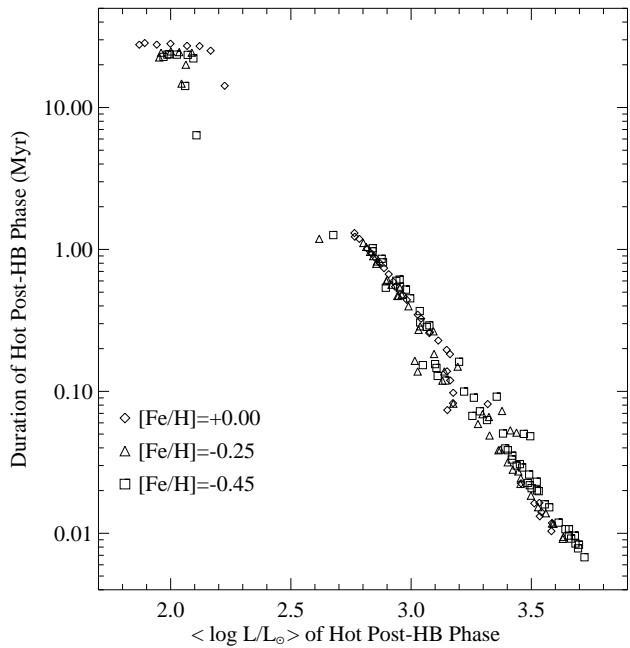} 
\epsscale{1.0}
\caption{The time spent as a hot post-HB star as a function of the
time-averaged luminosity in this phase, comparing our tracks at three
different metallicities.  The tight correlation shown in Figure~5 persists
even when the metallicity is varied.}  
\end{figure}

\begin{figure}[ht]
\epsscale{1.2}
\plotone{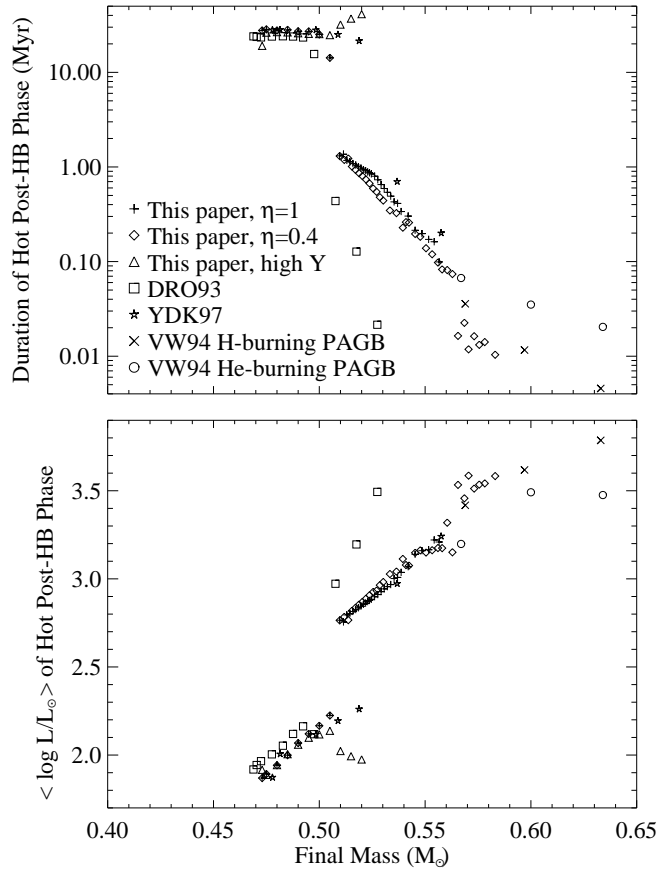} 
\epsscale{1.0}
\caption{{\it Top panel:} The time spent as a hot post-HB star as a function 
of the final mass on the WD cooling curve, shown for our own evolutionary
sequences and other models from the literature.  
{\it Bottom panel:} The time-averaged luminosity in the hot post-HB phase
as a function of the final mass on the WD cooling curve.  }
\end{figure}

\begin{figure}[ht]
\epsscale{1.2}
\plotone{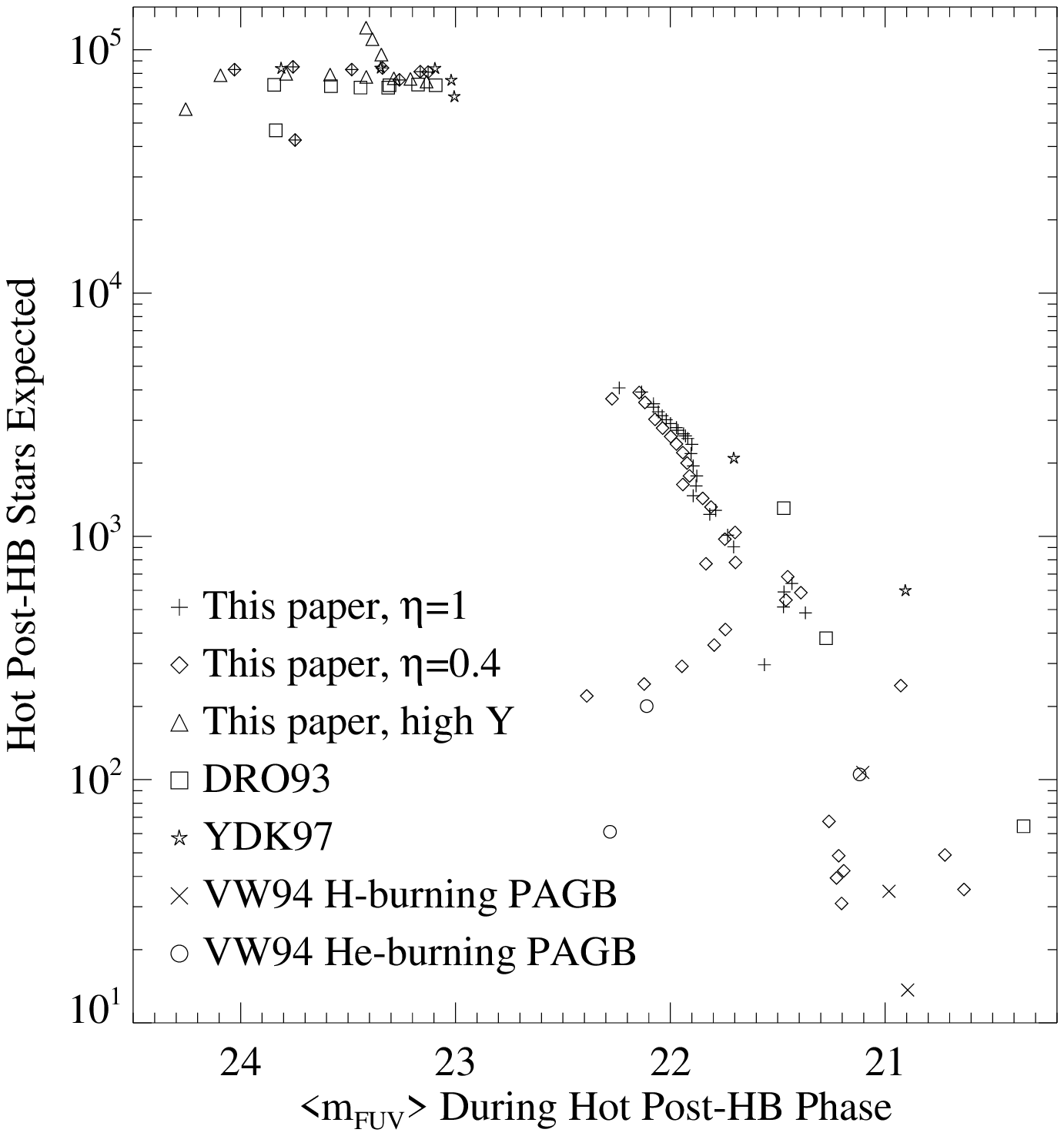} 
\epsscale{1.0}
\caption{The same as in Figure~5, but shown in terms of observable
parameters.  For each evolutionary track, the expected number of hot
post-HB stars assumes that all of the SEF in our observed region of
M32 evolves along that track.  It is thus an upper limit on the stars
that can be following that track, because in reality the SEF will be
distributed over a range of HB tracks.  Although there is still a clear
relationship between the number of expected stars and their mean far-UV
luminosity, the scatter is larger here than in Figure~5 due to the
large variation in bolometric correction across the hot post-HB region. }
\end{figure}

The number of stars evolving along any particular post-HB track is
constrained by the stellar evolutionary flux (SEF) of the population, i.e.,
by the rate at which the stars in the population are leaving any evolutionary
phase.  In general, the number $N_j$ of stars in any
particular evolutionary phase $j$ is the product of the duration of
that phase $t_j$, the specific evolutionary flux of the population
$B(t)$ (i.e., the stellar death rate per unit luminosity), and the total
bolometric luminosity $L_T$ of the population (Greggio \& Renzini
1990; Renzini 1998): \begin{equation}
N_j = B(t) L_T t_j. \end{equation} 
$B(t)$ depends
weakly upon age and metallicity and has a value of 2$\pm 0.2 \times 10^{-11}$
stars yr$^{-1}$ $L_\odot^{-1}$ for ages of 1--10 Gyr.  The
SEF is then given by
$SEF \equiv N_j / t_j \equiv B(t) L_T$.  The expected number
of hot post-HB stars in the population then follows from the
SEF and the duration $t_j$ of the hot post-HB phase.

Our UV catalog covers an area of approximately $21^{\prime\prime} \times
23^{\prime\prime}$, excluding a 9.4 arcsec$^2$ area in the center
of the galaxy.  Based on WFPC2/F555W images of this area, we find its
optical luminosity is $V= 9.97$~mag.  Given a foreground extinction of
$E(B-V) = 0.08$~mag (Schlegel et al.\ 1998) and a distance of 770~kpc
(Freedman \& Madore 1991), we obtain $M_V = -14.71$~mag.  If we assume an 
age of
8~Gyr for a population of nearly solar metallicity, the bolometric
correction is $-0.875$~mag (Worthey 1994), with an uncertainty of
$\sim$0.1~mag for ages of 5--12~Gyr.  The bolometric luminosity is
thus $1.36 \times 10^8$~$L_\odot$, which gives an SEF of $3.0 \times
10^{-3}$ star yr$^{-1}$.  

In Figure~8, we transform the theoretical parameters in Figure~5
into the observational
plane (number of stars vs.\ $<$$m_{FUV}$$>$).  The expected number of stars
evolving along each track has been normalized to the total number of stars
leaving the main sequence in our cataloged region ($3.0 \times
10^{-3}$ star yr$^{-1}$).  In other words, Figure~8 shows for each
evolutionary track the number of hot post-HB stars that would be
present in this region if the entire population evolved along that
particular track.  In reality, there is a dispersion in the HB mass
resulting from a range in mass loss on the RGB, which
leads to a dispersion in the post-HB behavior.  Thus the
total SEF should be distributed amongst the available tracks,
as we will do in our CMD simulations in \S4.

There is a clear progression from the long-lived AGBM tracks that
would yield $\sim$10$^5$ UV-bright stars in our field to the
short-lived PAGB tracks that would yield $\sim$10 UV-bright stars in
our field.  There is more scatter in Figure~8 than Figure~5, due to
the large bolometric correction at high effective temperature, even in
these UV bandpasses.

Our CMD simulations will include the effects of completeness and
photometric scatter, but putting those effects aside for the moment,
it is clear from Figure~8 that few of the stars in M32 are evolving
along AGBM tracks.  If every star leaving the main sequence evolved
along such tracks, our catalog would include $\sim$10$^5$ AGBM stars,
with approximately ten times as many EHB progenitors -- far more
AGBM and EHB stars than actually observed in our UV CMD of M32.
Such a large number of AGBM and EHB stars would also produce a
much stronger UV-upturn in M32 than observed.  Indeed, the observed
UV upturn in M32 requires that only $\approx$2\% of the HB stars lie on 
the EHB.  This
discrepancy implies that most of the population must evolve from the
red HB to become either PEAGB or PAGB stars, but there is a problem
with those tracks, too: the brightest PAGB tracks produce tens of
stars brighter than any seen in our image, while the fainter PAGB and
PEAGB tracks produce hundreds of stars at magnitudes where we detect
only a few such stars. 

We conclude that current stellar evolutionary models predict many more
hot post-HB stars in M32 than are actually observed regardless of how
these stars evolve from the HB to the WD cooling curve.  We will explore
this discrepancy further in \S4, where we will use CMD simulations, 
incorporating the photometric scatter and incompleteness 
from the data, to analyze the M32 UV CMD.

\subsection{Theoretical Luminosity Gap}

Virtually all of the helium-burning luminosity in an HB star is produced
within the innermost $\sim$0.04~$M_\odot$ of the core.  The
energy from this region
is carried outward by convection through the convective
core and then by radiation through the semiconvective zone and the
outer radiative region of the core.  With time the central helium
abundance ($Y_c$) decreases.  However, due to the high temperature dependence
of the helium-burning reactions ($\propto T^{35}$) an HB star is able to
compensate for this decrease in $Y_c$ by a small increase in the
central temperature.  Indeed the helium-burning luminosity actually
increases throughout the HB phase, as the helium
in the core is converted to carbon and oxygen and
the convective core and semiconvective zone grow in extent.  This
increase in the helium-burning
luminosity is responsible for the luminosity width of the EHB, where
helium burning is the dominant energy source.

\begin{figure}[ht]
\epsscale{1.2}
\plotone{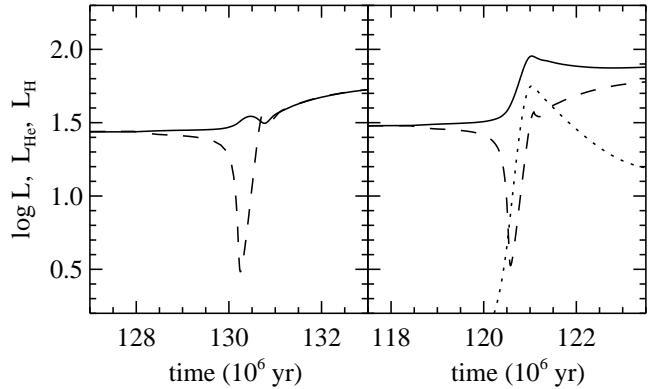} 
\epsscale{1.0}
\caption{Time dependence of the surface luminosity ({\it solid curve}), 
helium-burning
luminosity ({\it dashed curve}), 
and hydrogen-burning luminosity ({\it dotted curve}) at the
end of the HB phase for two EHB stars.  The left panel refers to a star at the
hot end of the EHB with a ZAHB effective temperature of 
log $T_{\rm eff}$ = 4.46,
while the right panel refers to a star at the cool end of the EHB with a ZAHB
effective temperature of log $T_{\rm eff}$ = 4.23.  The abscissa gives the time
elapsed since the ZAHB phase in units of $10^6$ yr.  All luminosities
are in solar units.}
\end{figure}

At the end of the HB phase, $Y_c$ goes to zero and helium burning ceases
at the center.  This leads to an abrupt drop in the helium-burning luminosity
and to the disappearance of the convective core.  In response to the
loss of this major energy source, the core of an HB star 
is forced to contract, as
the star tries to maintain its surface luminosity through the release
of gravitational potential energy.  This contraction heats the layers within
the former semiconvective zone until the temperature within these layers
rises sufficiently for helium burning to begin in a shell.  This sequence of
events is illustrated in the left panel of Figure~9, where we show
the time dependence of the surface luminosity and the helium-burning
luminosity at the end of the HB phase for a star at the hot end of the
EHB.  The drop in the helium-burning luminosity at the end of the HB phase
reaches a minimum at time t = 130.2 Myr.  Following
this minimum the helium-burning luminosity recovers as the newly formed 
helium-burning
shell stabilizes and the star returns to thermal equilibrium.
Note that the transition from central helium to helium-shell burning
is marked by an increase in the surface luminosity.  Hydrogen
burning during this phase is completely negligible
due to the very small envelope mass of this
star ($6 \times 10^{-4}$~$M_\odot$).

The same sequence of events occurs in cooler HB stars but with one important
difference.  The right panel of Figure~9 shows the time dependence of the
surface, helium-burning, and hydrogen-burning luminosities at the end of
the HB phase for a star near the red end of the EHB.  Again we see the
drop in the helium-burning luminosity associated with the transition
from central helium to helium-shell burning.  The contraction of the
core during this transition 
also raises the temperature within the hydrogen-burning
shell.  Due to the larger envelope mass ($\sim$0.02~$M_\odot$)
of the star in the right panel of Figure~9, the hydrogen-burning shell
turns on during this contraction and becomes a major energy source
for the star.  This leads to a sudden increase in the surface luminosity
which should appear observationally as a luminosity gap between the
EHB and post-EHB stars.  Such a luminosity gap is actually observed in 
globular clusters with cooler HB stars where the base of the AGB
appears as a clump well separated from the HB.  This AGB clump has
been detected in 47 Tuc (Hesser et al.\ 1987; Montegriffo et al.\ 1995)
as well as in a number of other clusters (Ferraro et al.\ 1999).  In a
study of the early AGB evolution in globular clusters, Cassisi et
al.\ (2001) found that the size of the predicted luminosity gap between
the HB and AGB is not affected by the use of updated input physics in
their stellar models.

The properties of this luminosity gap are illustrated in the top panel
of Figure~10 for 6 EHB and post-EHB evolutionary sequences with $Z$ = 0.01
and $Y$ = 0.23.  We plot each track in this panel by a
series of points spaced every $10^6$ yr during the evolution in order to show
where one would expect to find stars in an actual CMD.
The hottest track in this panel lies at
the hot end of the EHB with a ZAHB effective temperature
of log $T_{\rm eff}$ = 4.46.  The masses of the cooler EHB tracks
increase with an increment of 0.005~$M_\odot$.  The tracks in this
panel cover the mass range that yields hot post-HB stars.  The
luminosity gap predicted by our models can be clearly seen
between the EHB and post-EHB phases.

The size of the luminosity gap depends on how strongly the hydrogen-burning
shell turns on at the end of the HB phase and hence on the envelope
mass of the star.  Because a helium-rich EHB star has a larger envelope
mass at a given effective temperature, one would expect a larger
luminosity gap at higher helium abundances.  This expectation is
confirmed in the middle and bottom panels of Figure~10, where we plot
the EHB and post-EHB tracks for the same value of $Z$ ($=0.01$) but for 
$Y$ values of 0.28 and 0.33.
As in the top panel, each track is again plotted as a
series of points spaced every $10^6$ yr during the evolution, and
the mass increment between successive tracks is again 0.005~$M_\odot$.
Besides the larger luminosity gap for the helium-rich compositions,
we also note that the mass range of EHB stars that evolve into
hot post-EHB stars is considerably larger than in the top panel
of Figure~10.

The luminosity gap shown in Figure~10 is a consequence of the
changes that occur in
the basic structure of a star at the end of the EHB phase 
and therefore should be a robust prediction of the theoretical models.
In the following section we will use CMD simulations derived from our
evolutionary tracks to compare this
predicted luminosity gap with the observed UV CMD of M32.

\begin{figure}[ht]
\epsscale{1.25}
\plotone{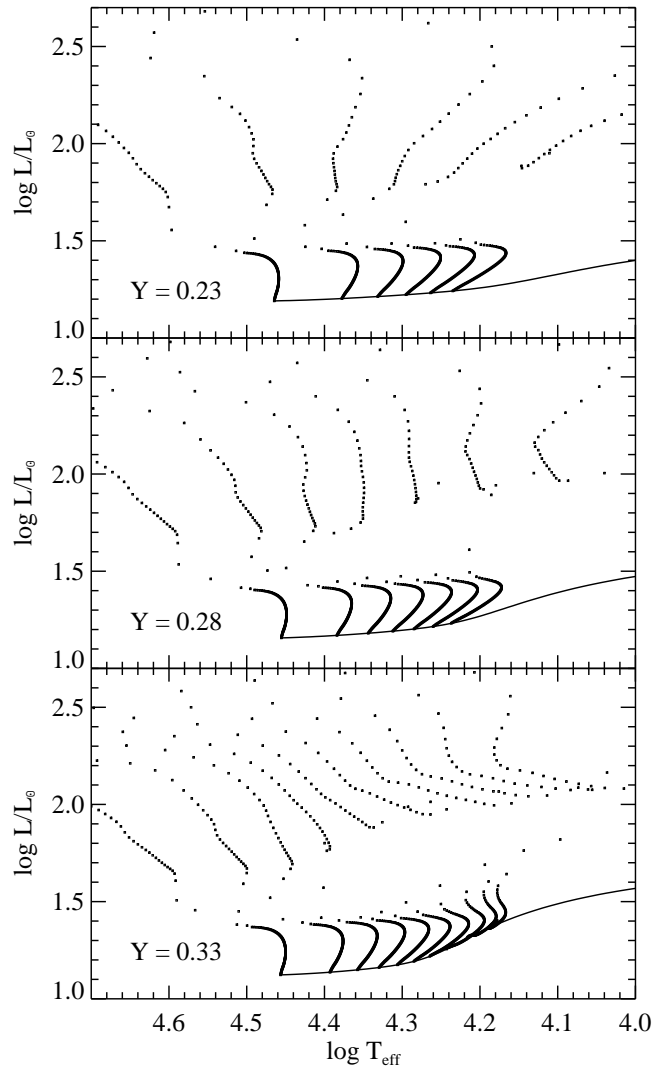} 
\epsscale{1.0}
\caption{EHB and post-EHB evolutionary tracks for $Z$ = 0.01 and $Y$ = 0.23
({\it top panel}), 0.28 ({\it middle panel}), and 0.33 ({\it bottom panel}).  
Each track is
plotted by a series of points spaced every $10^6$ yr during the evolution.
The hottest track in each panel lies at the hot end of the EHB.  The mass
spacing between the tracks in each panel is 0.005~$M_\odot$.  The solid
curves represent the ZAHB for each composition.}
\end{figure}

\begin{figure*}[ht]
\epsscale{1.0}
\plotone{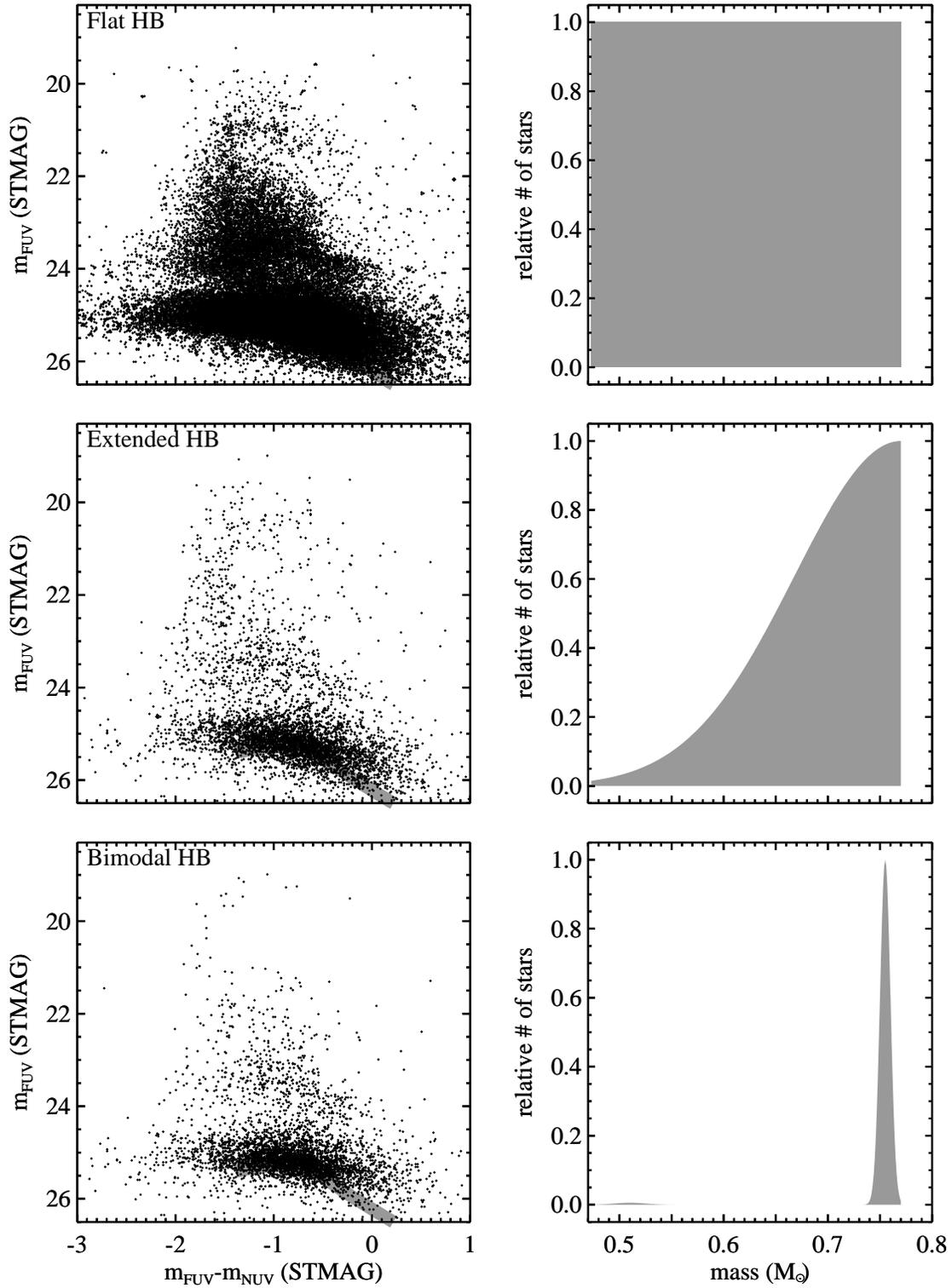} 
\epsscale{1.0}
\caption{Simulated CMDs ({\it left panels}) corresponding to
arbitrary distributions of mass on the HB ({\it right panels}).
A flat distribution of mass on the HB ({\it top panels}) produces
far too many stars on and above the EHB.  A predominantly red HB
with a Gaussian extension to the blue ({\it middle panels}) can be tuned
to approximate the observed number of EHB stars, but predicts far too many
UV-bright stars above the EHB.  A bimodal distribution of HB mass
({\it bottom panels}) can reduce the predicted number of UV-bright stars above
the EHB, but it is still difficult to reproduce both the observed number of
these stars and their luminosity (see text).  Also, the gap between
the EHB and the AGBM is much more pronounced in the simulations
than in the data.
}
\end{figure*}

\begin{figure}[ht]
\epsscale{1.2}
\plotone{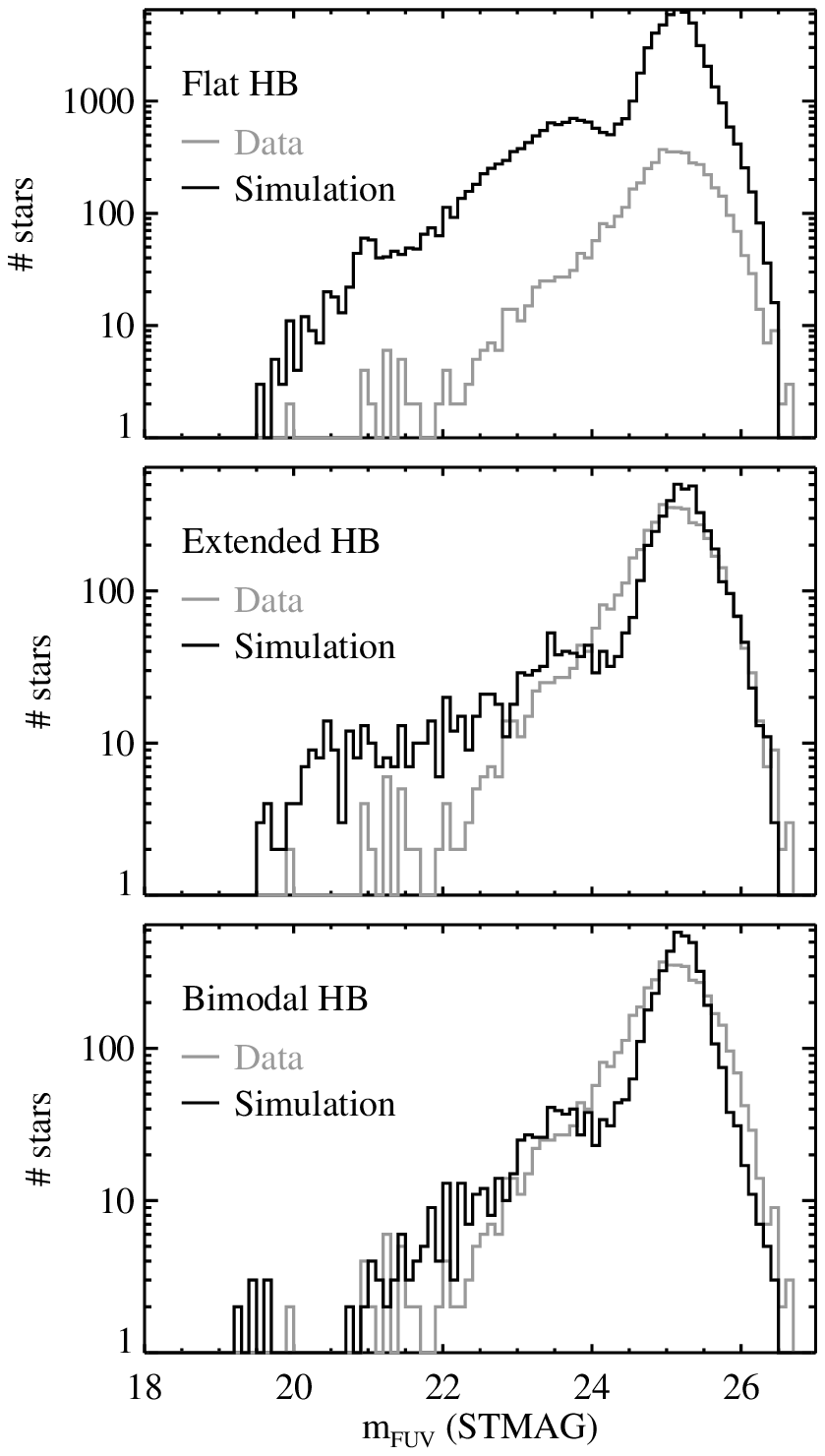} 
\epsscale{1.0}
\caption{Luminosity functions for each of the simulated CMDs
from Figure~11 ({\it black histograms}) compared to the observed far-UV
luminosity function ({\it grey histogram}).  The luminosity function
spans the color range $-2 \leq (m_{FUV}-m_{NUV}) \leq 0$~mag.  Instead of 
correcting the observed luminosity function for completeness, the
model luminosity functions include both the completeness and photometric
scatter, determined from artificial star tests.  The luminosity gap
between the EHB and AGBM stars is very clear in the simulated
CMDs, but not obviously present in the observed population.
}
\end{figure}

\section{CMD Simulations}

\subsection{Number of UV-Bright Stars}

Our initial simulations are simple models that demonstrate the issues
discussed in previous sections.  At times we will follow these fits with
standard numerical algorithms for CMD fitting, but the utility of such
techniques is hampered by the fundamental discrepancies between the models
and the data.  We start with solar-abundance models
at a distance of 770~kpc (Freedman \& Madore 1990) and an extinction
of $E(B-V) = 0.08$~mag.  First we vary the distribution of mass on the
HB.  In Figure~11, we show three different distributions of mass: a
flat distribution, a red HB with an extension to the blue, and a
bimodal HB (two Gaussians).  In each simulation, the total number of HB stars
is consistent with the SEF of $3.0 \times 10^{-3}$
stars yr$^{-1}$, but the number of stars that appear in the UV varies
dramatically.  The simulated CMDs are shown in the left-hand panels,
while the corresponding mass distributions are shown in the right-hand
panels.  In Figure~12, we show a comparison
between the observed far-UV luminosity function and the luminosity
function derived for each of the simulated CMDs.

The flat distribution ({\it top panels} in Figures 11 and 12) 
obviously does not work -- it
is included only as a starting point for the discussion.  Note that
this is a flat distribution in mass, not a flat distribution in
effective temperature -- the bulk of the HB stars in this distribution
are on the red HB.  Far too many stars are populating the EHB and
post-HB phases.  To reproduce the observed CMD, we clearly need only a
small minority of stars at low envelope mass.

The extended HB does much better, but is still problematic.  This
simulation was created using a Gaussian distribution of mass centered
on the red end of the HB, with the dispersion in mass tuned to
reproduce the number of EHB stars observed in our M31 CMD, with
$\sim$2\% of the HB stars lying on the EHB.  However, there are far
too many UV-bright stars above the HB, compared to observations.  Thus
there are too many stars on those PEAGB and PAGB tracks that are
relatively long-lived, with intermediate HB masses (between those with
the lowest mass, which evolve into long-lived AGBM stars, and those
with the highest mass, which evolve into very short-lived PAGB stars).

The bimodal HB does even better, by transferring weight from the
PEAGB-producing tracks to the PAGB-producing tracks, thus reducing the
number of UV-bright stars above the EHB, but there are still several
discrepancies.  There are more PAGB stars than observed, and they
extend to brighter luminosities.  Furthermore, this simulation predicts a
pronounced luminosity gap between the EHB and post-HB stars, which
does not appear in the observed population.  The luminosity
width of the simulated HB is also much narrower than the observed HB.

We next attempt to improve the fit to the
observed CMD by letting the weights of the EHB tracks float freely (instead
of fixing them as in the bimodal Gaussian distribution of Figure 11).
Specifically, we fit the data via minimization of a Maximum Likelihood
statistic, using an amoeba algorithm to drive the weights of each
evolutionary track, and binning the data and models by 0.05~mag in
both $m_{FUV}-m_{NUV}$ color and $m_{FUV}$ luminosity.  In this
fitting, we enforced the constraint that the total weight in all
tracks must equal the SEF of $3 \times 10^{-3}$ stars yr$^{-1}$.  The
weights for the tracks on the EHB were allowed to float freely within
this constraint, and any remaining weight was put into the reddest HB
track, which minimizes the number of PAGB stars.  
The resulting fit is still a bimodal mass distribution,
but this is only because we are excluding those intermediate-mass
tracks that lead to PEAGB behavior, given that they produce far too
many UV-bright stars.  The resulting fit, given in
Figure~13, shows only slightly better agreement with the data.  When
there are fundamental discrepancies between the models and data, it is
not very meaningful to quantify the goodness of fit, but we do so here
for completeness.  Our fit minimized a Maximum Likelihood statistic
instead of $\chi^2$, so we provide an effective $\chi^2$ ($\chi^2_{\rm
eff}$; Dolphin 2002) for readers more familiar with such fitting:
$\chi^2_{\rm eff}$~=~1.6.  The fit is strongly ruled out at
38$\sigma$.

\begin{figure}[ht]
\epsscale{1.2}
\plotone{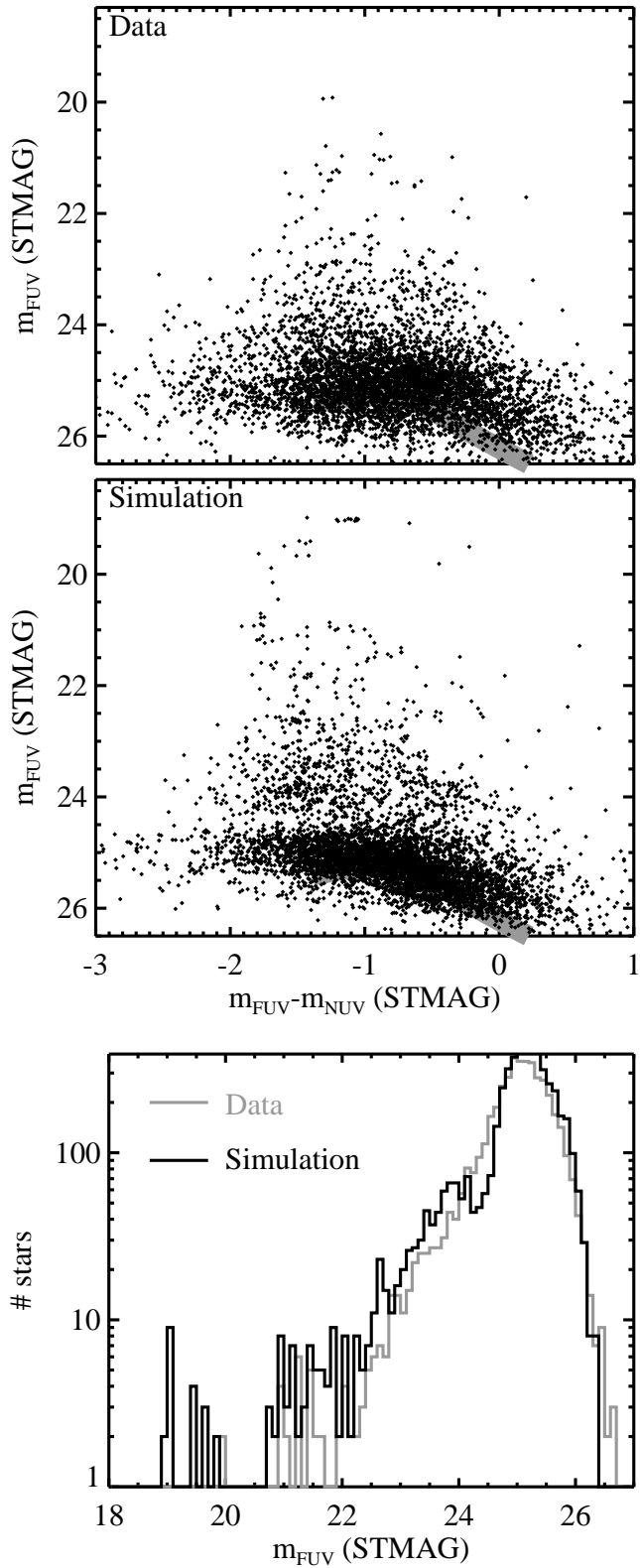} 
\epsscale{1.0}
\caption{{\it Top:} The observed far-UV CMD of M32.  {\it Middle:} A
simulated CMD arising from a bimodal distribution of mass on
the HB, where the weights of the EHB tracks are determined by a 
Maximum Likelihood statistic, and
the remainder of the population falls on the reddest HB track. 
{\it Bottom:} The comparison of
the observed and simulated luminosity functions for those stars
at $-2 \leq (m_{FUV}-m_{NUV}) \leq 0$~mag.
\vspace{0.2in}
}
\end{figure}

Using other PAGB models in the literature, such as those of VW94, does
not produce better agreement.  In Figure~14, we take the 
bimodal mass distribution of Figure~11 to populate the EHB, but 
evolve the red HB stars 
along one of the VW94 PAGB tracks instead of our own.  In Figure 14, the
left-hand panels show the H-burning tracks of VW94, while the right-hand
panels show their He-burning tracks.  CMD simulations are shown 
for main sequence
masses of 1, 1.5, and 2~$M_\odot$ for both types of PAGB tracks.  
The corresponding luminosity functions
are shown in Figure~15.  The same PAGB problem persists: for low-mass PAGB
tracks,
too many UV-bright stars appear at the bright end of our observed range,
while for high-mass PAGB tracks, the UV-bright stars extend to magnitudes much
brighter than observed.  This result should not be surprising, given
the agreement between our PAGB tracks and those of VW94 in Figures 5 and 8. 
Thus for all the models in hand, the discrepancies between the predicted and 
observed distributions of PAGB stars seem unavoidable.

\begin{figure*}[ht]
\epsscale{1.0}
\plotone{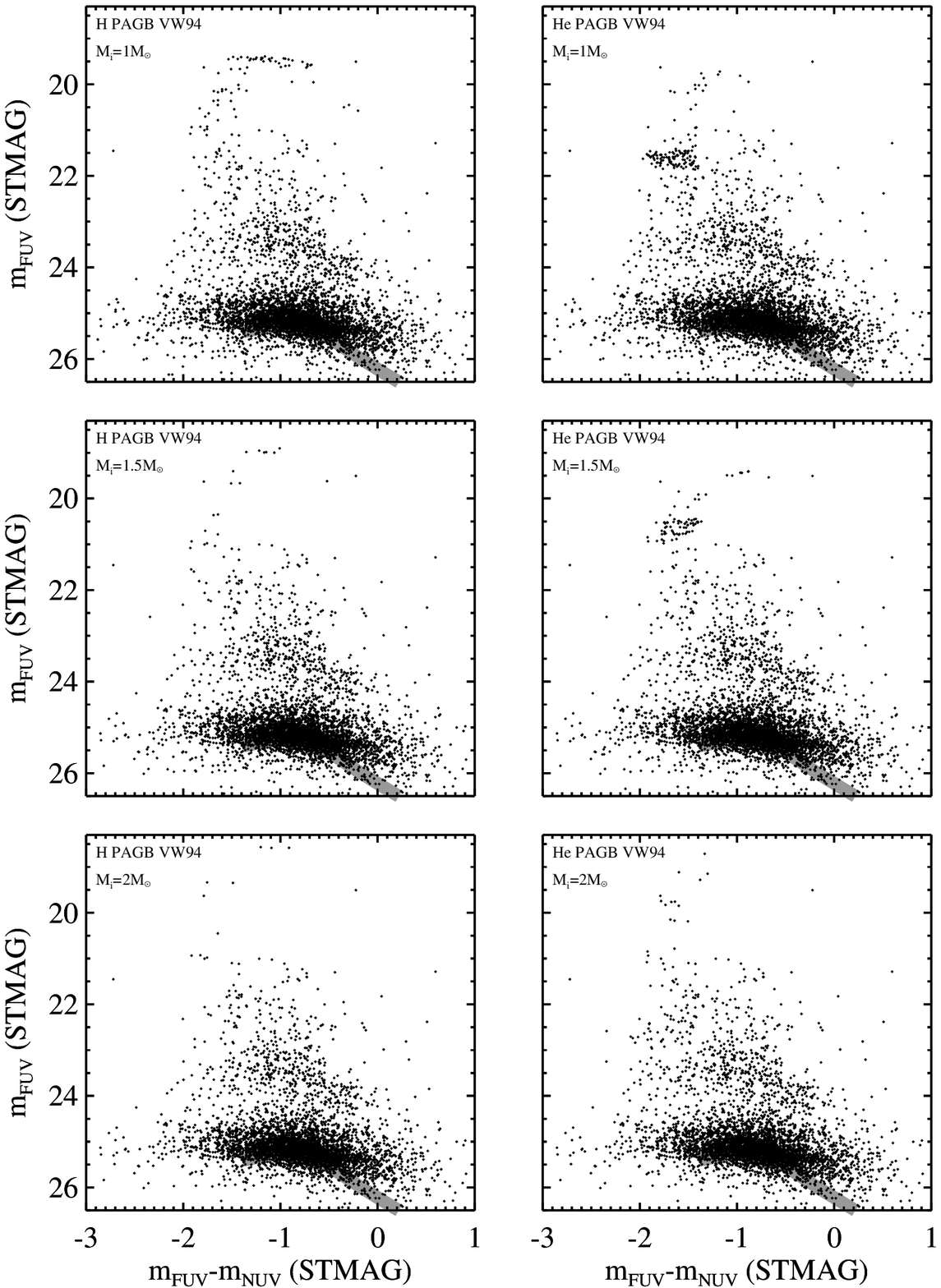} 
\epsscale{1.0}
\caption{Simulated CMDs corresponding to a bimodal
distribution of mass on the HB, as given in Figure~11, but with the
red HB stars evolving along the PAGB tracks of VW94 instead of our own.
Each panel is labeled with the main sequence mass and the PAGB class.}
\end{figure*}

\begin{figure*}[ht]
\epsscale{1.2}
\plotone{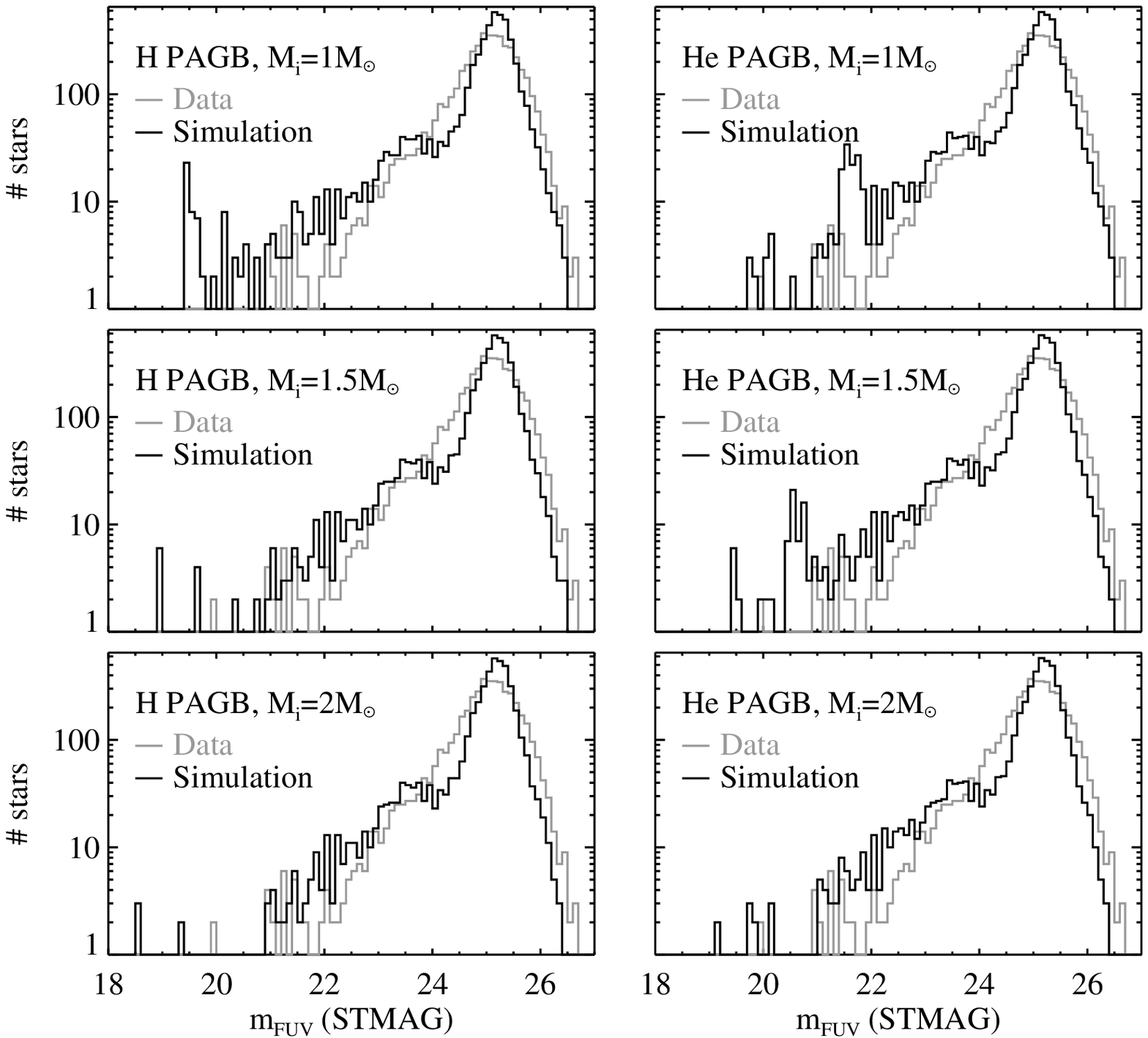} 
\epsscale{1.0}
\caption{Luminosity functions for each of the simulated CMDs
from Figure~14 ({\it black histograms}) compared to the observed far-UV
luminosity function ({\it grey histograms}).  Depending upon the assumed mass
of the PAGB stars, the simulations produce PAGB stars far brighter than
observed, and/or too many PAGB stars near the bright end of our observed
range of luminosities.
}
\end{figure*}

\subsection{Luminosity Gap}

We now turn our attention to the luminosity gap between the EHB and
post-HB phases.  This gap is very pronounced in the simulations so
far, in contrast to the observations.  There are several 
parameters that can affect the distribution of HB and post-HB stars
and consequently the size of the predicted luminosity gap,
including metallicity, helium abundance, and extinction.  We will
explore each of these in turn.

First we consider metallicity.  In Figure~16, we show simulations
assuming the same bimodal distribution of mass given in Figure~11, but
for [Fe/H] values of $-0.25$ and $-0.45$ instead of solar.  The
corresponding luminosity functions are shown in Figure~17.  We next
combine [Fe/H] values of 0, $-0.25$, and $-0.45$ with relative weights
of 25\%, 50\%, and 25\%, respectively, to approximate the metallicity
distribution of Worthey et al.\ (2004).  The resulting simulation (given
in the lower left-hand panel of Figure~16)
shows some blurring of the luminosity gap, but not enough to account
for the observations.  Thus, neither a combination of metallicities
nor a single metallicity will account for the absence of a luminosity gap.

Next, we explore the effects of enhancing the helium abundance.  With
all other parameters (age, metallicity, mass loss, etc.) held fixed, a
population will form more EHB stars as the helium abundance is
increased (Greggio \& Renzini 1990).  This is because an enhanced helium
abundance reduces the 
mass at the main sequence turnoff, thus leading to HB stars of lower mass,
and because helium-rich EHB stars of a given envelope mass have a higher
effective temperature (see Figure~10).  High
values of $\Delta Y / \Delta Z$ are thus one of the possible
explanations for producing EHB stars in relatively metal-rich
elliptical galaxies (e.g., O'Connell et al.\ 1999; YDK97; DRO93).
In Figure~16, we show a simulation with the same bimodal
mass distribution employed in the previous simulations, but with $Y$
values of 0.28 and 0.43 for the EHB stars,
corresponding to $\Delta Y / \Delta Z = 4$ and 19, respectively.
The corresponding luminosity functions are shown in Figure~17.
Increasing the helium abundance has two notable effects on the UV CMD:
it increases the size of the luminosity gap between the EHB and AGBM
phases, and it increases the size of the AGBM population relative to
the EHB.  Both of these effects serve only
to increase the discrepancy between the models and the observed UV CMD.  This
$Y$ sensitivity of the UV CMD was one of the main motivations for our
observations, given the difficulty in measuring $Y$ in elliptical
galaxies, but our UV CMD is apparently inconsistent with an enhanced
$Y$.

Finally, we explore the effects of extinction.  Because the effects of
extinction are so much stronger in the UV than in the optical, a
relatively small dispersion in $E(B-V)$ can produce a large spread in
the UV.  As an example, 
we assume a Gaussian distribution of $E(B-V)$ values, with a
mean value of 0.08~mag (Schlegel et al.\ 1998), a one-sigma width of
0.04~mag, and the mean Galactic extinction law (Fitzpatrick 1999); the
Gaussian was truncated to avoid negative (unphysical) values of
$E(B-V)$.  The resulting CMD is shown in Figure~16, and the luminosity
function is shown in Figure~17.  This small dispersion in extinction
does produce much better agreement with the observations.  In particular, 
the luminosity gap in the simulation is much less pronounced.

\begin{figure*}[ht]
\epsscale{1.0}
\plotone{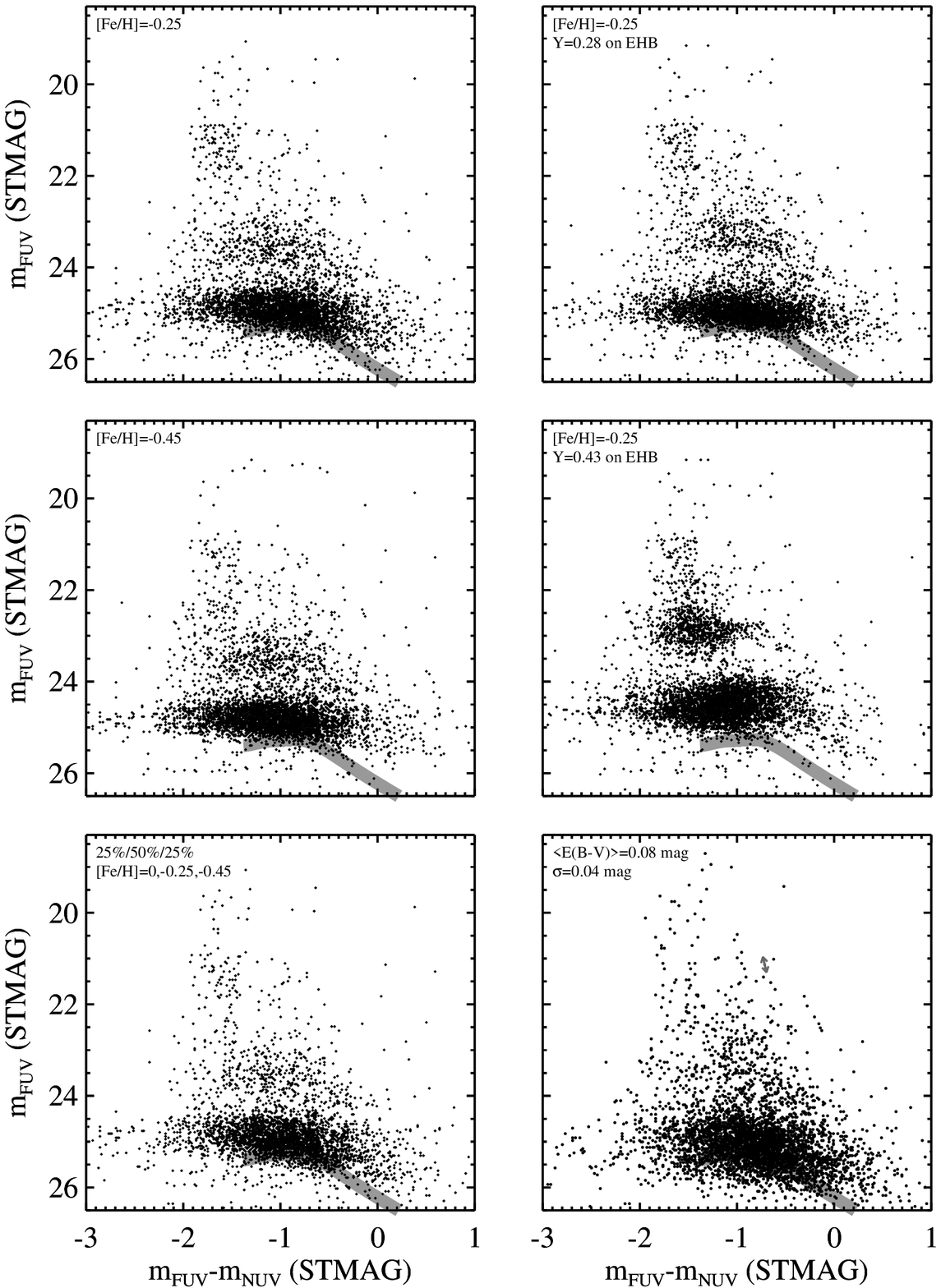} 
\epsscale{1.0}
\caption{Simulated CMDs corresponding to a bimodal distribution of
mass on the HB, as in Figure~11, but with variations in [Fe/H], $Y$,
and extinction.  A distribution of [Fe/H] ({\it lower left-hand
panel}) does not blur the luminosity gap enough to match the observed
CMD, but a Gaussian distribution of $E(B-V)$ ({\it lower right-hand
panel}) does.  The extinction vector is shown in grey in the lower left-hand
panel, scaled to the CMD displacement arising from 
$E(B-V) = 0.08 \pm 0.02$~mag.
Enhancing the helium abundance for the EHB
population causes a more prominent luminosity gap, which is very
discrepant with the observations.}
\end{figure*}

\begin{figure*}[ht]
\epsscale{1.2}
\plotone{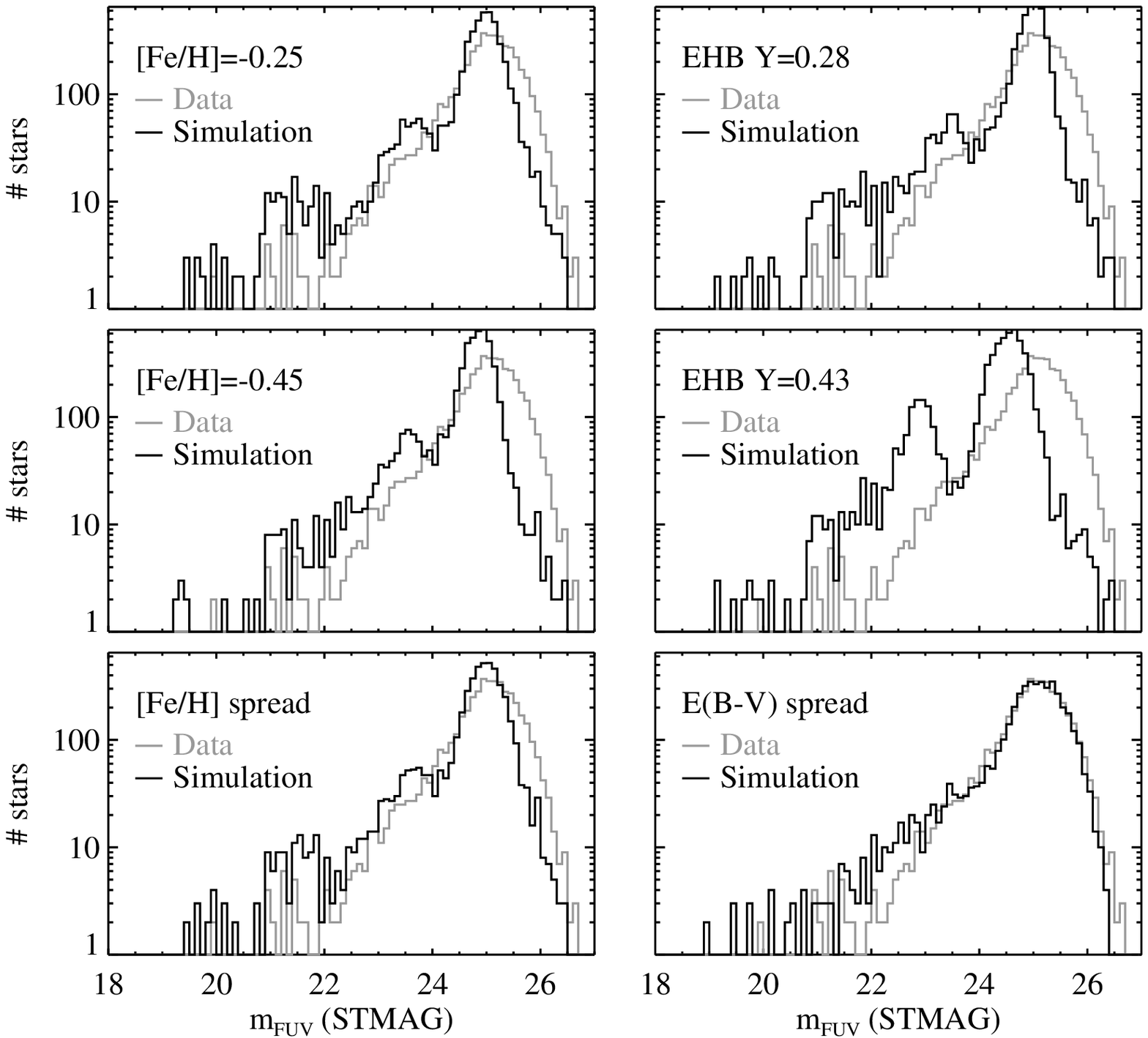} 
\epsscale{1.0}
\caption{Luminosity functions for each of the simulated CMDs
from Figure~16 ({\it black histograms}) compared to the observed far-UV
luminosity function ({\it grey histograms}).
}
\end{figure*}

Given the promising results of the simulation with an extinction
spread, we decided to improve 
this simulation by replacing the arbitrary bimodal
mass distribution used in Figure 11 with a fit to the observed CMD,
based on the minimization of a Maximum Likelihood statistic, as in Figure 13.
Again, we enforced the constraint that the total number of stars on
all tracks must be consistent with 
the SEF of $3 \times 10^{-3}$ stars yr$^{-1}$.
The weights for the individual EHB tracks were allowed to float freely
within this constraint, and any remaining weight was put into the
reddest HB.  All other parameters were held fixed:
[Fe/H]=$-0.25$ (the peak in the M32 metallicity distribution; Worthey et
al.\ 2004), solar $Y$, and a Gaussian distribution of $E(B-V)$ with a
mean of 0.08 mag and a one-sigma width of 0.04 mag.  The result is
shown in Figure~18, which compares the CMD and luminosity function in the
data and best fit.  Although this fit is much better than our earlier
attempts ($\chi^2_{\rm eff}$~=~1.2), 
there are still clear discrepancies, and the 
fit is formally ruled out at 15$\sigma$.  In the
simulation, the UV-bright post-HB population includes too many stars
that extend to brighter magnitudes than observed, but as discussed previously,
there is no way
to address this problem with the models at our disposal.  Furthermore,
the EHB is more sloped in the simulation than in the observed CMD;
this may be due to the particular distribution of age, [Fe/H], $Y$,
and mass in the observed population (including correlations between
these parameters), but given the number of free parameters and the
limitations of these models, we will not try to resolve this
discrepancy here.

\section{Discussion}

We have presented a UV CMD for a field in the center of the compact
elliptical galaxy M32.  As found in our previous analysis of the
near-UV data alone, we find that the UV light in M32 is clearly due to
a minority ($\sim 2$\%) population of EHB stars, and that there is a
dearth of UV-bright post-HB stars (AGBM, PEAGB, and PAGB stars) when
comparing the observed CMD to the expectations of stellar evolution
models.  Although the presence of EHB stars in metal-rich elliptical
galaxies has often been attributed to the possible enhancement of
helium at high metallicity (see O'Connell 1999 and references
therein), our UV CMD is inconsistent with a super-solar $Y$ in the EHB
population.  Compared to the models, we find a wider far-UV luminosity spread
in the EHB, and little (if any) luminosity gap between the EHB and AGBM
phases.

The inconsistency between the models and data can be largely rectified
by assuming a small dispersion ($\sigma \approx 0.04$~mag) in $E(B-V)$
around the mean value of 0.08~mag.  This dispersion in extinction
gives our simulated EHB a luminosity width nearly identical to
that observed, but the agreement with the observed CMD is still not
perfect, due to a stronger tilt in the simulated EHB, and to the larger
predicted number of UV-bright stars (see previous section).  

If we are
really detecting a dispersion in extinction along the sightlines to
M32's evolved stars, it is not clear where this extinction lies.  We
see no evidence of large-scale extinction patchiness in our field
(comparing the CMD of distinct regions shows no shift in mean EHB
luminosity or EHB width).  A variation in foreground Galactic
extinction seems improbable, given the fraction of the Galaxy
intercepted by the small STIS field.  The M31 disk would be a better
candidate for variable extinction, but it is unclear if M32 lies
behind the M31 disk or in front of it (Worthey et al.\ 2004; Ford et
al.\ 1978; Paulin-Henriksson 2002).  The variable extinction could be
due to a small amount of dust in the center of M32, but observations
suggest there is not much (Corbin et al.\ 2001).  The varying amount
of dust could conceivably be localized to these evolved stars, given
the mass lost in their earlier evolutionary phases.  Our far-UV
bandpass falls to the blue of the 220 nm extinction bump, while our
near-UV bandpass falls on top of this bump, so appropriately deep
observations near the $U$-band would be useful in explorations of this
extinction possibility.

As discussed by Brown et al.\ (2000b), there are several possible
explanations for the lack of PAGB stars in our field.  The majority of
the PAGB stars might be crossing the HR diagram on a thermal timescale
instead of a nuclear timescale; this could happen if the final mass
ejection on the AGB is triggered by a helium-shell flash that leaves
the star out of thermal equilibrium (K$\ddot{\rm a}$ufl et al 1993;
Greggio \& Renzini 1999).  PAGB stars will also cross the HR diagram more
rapidly if they are more massive, but this is problematic, because
such stars will cross at luminosities much brighter than we observe;
such luminosities would also imply an AGB tip luminosity much brighter
than observed (Greggio \& Renzini 1999).  Another possibility is that
the PAGB stars are obscured by circumstellar material, but this solution
is also problematic, given the rapid thinning times for material surrounding
post-AGB stars (K$\ddot{\rm a}$ufl et al. 1993).

Compared to other elliptical galaxies, M32 has an extremely weak UV
upturn (Burstein et al.\ 1988).  In principle, all of its UV emission
could arise from the PAGB descendents of an entirely red HB population.
Instead, we find that nearly all of its UV
emission is due to a minority population of EHB stars and their AGBM
progeny.  Most of the HB stars reside on the red end of the HB, but 
only a small fraction of their expected PAGB descendents are present.  These
findings have significant implications for studies trying to use the
UV upturn as a tracer of age in old populations.  The UV upturn has great
potential as a tracer of age in a population, because as a population
ages, the mass at the main sequence turnoff decreases, leading to HB
stars of lower mass and hotter temperatures (if RGB mass loss is held
fixed).  For this reason, several studies have tried to map the
strength of the UV upturn in elliptical galaxies as a function of
redshift (Brown et al.\ 1998; Brown et al.\ 2000a; Brown et al.\ 2003;
Lee et al.\ 2005; Ree et al.\ 2007).  Unfortunately, at the present
time, the many parameters that govern the presence of EHB stars in a
population are poorly constrained; one can tune the assumptions
(chemical evolution, RGB mass loss, binaries, etc.) so that the sudden onset 
of the UV upturn in a population can occur at nearly any age greater than
a few Gyr.  Some models (e.g., those in Ree et al.\ 2007) assume that
EHB stars fade rapidly beyond $z \sim 0.1$, such that the UV emission
from populations at $z \gtrsim 0.3$ should be completely dominated by
PAGB stars.  The results we have presented here show that this is not
necessarily the case.  We do not know the age distribution in M32, but
at least part of the population is old enough to host EHB stars.  M32
is an elliptical galaxy at $z = 0$ with far fewer PAGB stars than one
would expect.  If we observed M32 in the past, it would have even
fewer PAGB stars, because the PAGB stars would be more massive and
evolve more rapidly.  The population of M32 implies that PAGB stars
are not a significant source of UV emission, and that at any redshift
where significant UV emission is present in a population, it is likely
due to either EHB stars or residual young populations; this is true
even for redshifts at the high end of the range explored in UV upturn
studies ($z$=0.55; Brown et al.\ 2000a).  Models that
predict a rapid fading of UV emission with increasing redshift should likely
be revised, such that the UV emission fades even more dramatically, because
the PAGB contribution to such emission is likely overestimated.

\begin{figure}[ht]
\epsscale{1.2}
\plotone{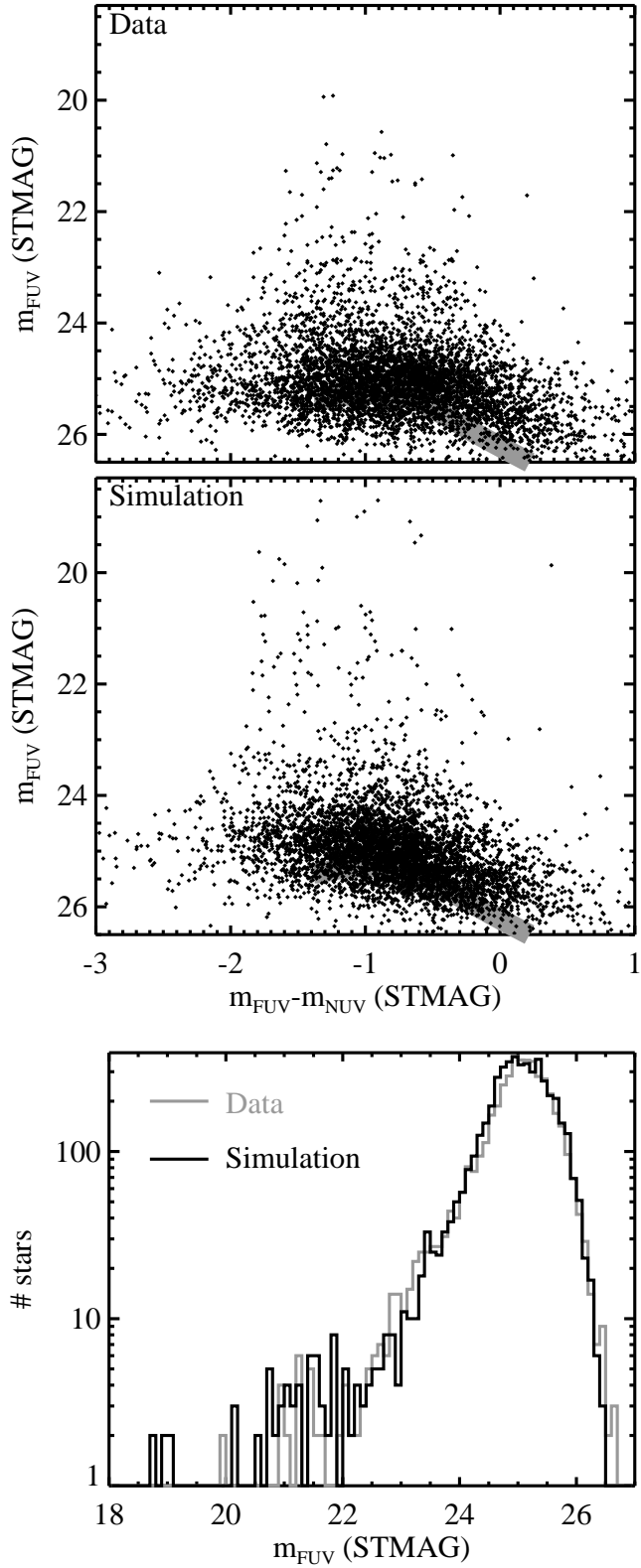} 
\epsscale{1.0}
\caption{{\it Top:} The observed far-UV CMD of M32.  {\it Middle:} A
simulated CMD arising from a bimodal distribution of mass on
the HB, where the weights of the EHB tracks are determined from a fit using
a Maximum Likelihood statistic, and
the remainder of the population falls on the reddest HB track. 
The [Fe/H] is fixed at $-0.25$, the $Y$ is fixed at the solar value,
and there is a Gaussian distribution of $E(B-V)$ with a mean of 0.08~mag
and a one-sigma width of 0.04~mag.  {\it Bottom:} The comparison of
the observed and simulated luminosity functions.}
\end{figure}

\acknowledgements

Support for proposal 9053 is provided by NASA through a grant from
STScI, which is operated by AURA, Inc., under NASA contract NAS
5-26555.  We are grateful to P.\ Stetson for his DAOPHOT
code, and to A.\ Renzini for interesting discussions.

\end{document}